\documentclass[aps,prx,twocolumn,floatfix,superscriptaddress,nofootinbib]{revtex4-1}
\usepackage{tabularx} 
\usepackage{amsmath}  
\usepackage{amssymb}
\usepackage{graphicx} 
\usepackage[final]{hyperref} 
\usepackage{multirow}
\hypersetup{
	colorlinks=true,       
	linkcolor=blue,        
	citecolor=blue,        
	filecolor=magenta,     
	urlcolor=blue         
}
\usepackage{blindtext}
\usepackage{verbatim}
\usepackage{siunitx}
\usepackage[normalem]{ulem}

\emergencystretch=\maxdimen
\newcommand{\Fig}[1]{Fig.~\ref{#1}}

\newcommand{\Eqs}[2]{Eqs.~(\ref{#1}) and (\ref{#2})}

\newcommand{\bvec}[1]{\mathbf{#1}}
\newcommand{\xhat}{\hat{\bvec{x}}}
\newcommand{\yhat}{\hat{\bvec{y}}}
\newcommand{\zhat}{\hat{\bvec{z}}}
\newcommand{\mhat}{\hat{\bvec{m}}}
\newcommand{\phat}{\hat{\bvec{p}}}
\newcommand{\torque}{\boldsymbol{\tau}}

\newcommand{\js}{j_s}
\newcommand{\jSHE}{j_s^\text{SHE}}
\newcommand{\jOOP}{j_s^\text{OOP}}
\newcommand{\ratio}{\beta}

\begin{document}

\title{Large-Amplitude, Easy-Plane Spin-Orbit Torque Oscillators Driven by Out-of-Plane Spin Current: A Micromagnetic Study}

\author{Daniel Kubler}
\affiliation{Department of Physics, Indiana University, Indianapolis, IN 46202, USA}

\author{David A. Smith}
\affiliation{Department of Physics, Virginia Tech, Blacksburg, VA 24061, USA}
\affiliation{HRL Laboratories, Malibu, CA 90265}

\author{Tommy Nguyen}
\affiliation{Department of Physics, Indiana University, Indianapolis, IN 46202, USA}

\author{Fernando Ramos-Diaz}
\affiliation{Department of Physics, Virginia Tech, Blacksburg, VA 24061, USA}

\author{Satoru Emori}
\email{semori@vt.edu}
\affiliation{Department of Physics, Virginia Tech, Blacksburg, VA 24061, USA}

\author{Vivek P. Amin}
\email{vpamin@iu.edu}
\affiliation{Department of Physics, Indiana University, Indianapolis, IN 46202, USA}

\date{\today}

\begin{abstract}

Spin torque oscillators are spintronic devices that generate a periodic output signal from a non-periodic input, making them promising candidates for applications like microwave communications and neuromorphic computing. However, traditional spin torque oscillators suffer from a limited precessional cone angle and thermal stability, as well as a need for an applied bias magnetic field. We use micromagnetic simulations to demonstrate a novel spin torque oscillator that relies on spin-orbit effects in ferromagnets to overcome these limitations. The key mechanism behind this oscillator is the generation of an \emph{out-of-plane spin current}, in which both the spin flow and the spin orientation are out-of-plane. The torque from this spin current enables easy-plane coherent magnetic precession with a large cone angle and high thermal stability over a micron-scale lateral area. Moreover, the precession occurs about an internal field in the free layer, thereby eliminating the need for an external bias field. We demonstrate the feasibility of an easy-plane spin-orbit torque oscillator at room temperature over a wide parameter space, including the ratio of the out-of-plane spin current to the conventional spin-Hall spin current, presenting exciting possibilities for this novel spintronic device.
\end{abstract}

\maketitle

\section{Introduction}
Devices that efficiently convert a dc input into a self-sustaining ac signal are crucial to a wide variety of fields from microwave communications to neuromorphic computing \cite{Locatelli_15,Grollier_16,Lim_20}. Spin torque oscillators are promising building blocks for such applications due to their GHz oscillation frequencies and purported energy efficiency \cite{Houssameddine2007,Troncoso_17,Rippard,Kiselev2003}. 
In any spin-torque oscillator, the magnetic order parameter of the magnetic free layer self-oscillates under a dc electrical current input. In particular, the self-sustained oscillations are driven by a spin torque, i.e.,~a transfer of spin angular momentum from an incident spin-polarized current to the magnetization of a ferromagnetic layer \cite{SLONCZEWSKI_5,Stiles_6}. The spin torque effectively cancels magnetic damping in the free layer -- thereby allowing the magnetization to precess freely about a magnetic field. The precessing magnetization generates an oscillating electrical voltage output, i.e., the product of the dc current and the time-varying magnetoresistance. For high power output from a spin torque oscillator, it is critical to stabilize a large cone angle for magnetic precession. 

\begin{figure*}[] 
   \centering
   \includegraphics[width=0.875\textwidth]{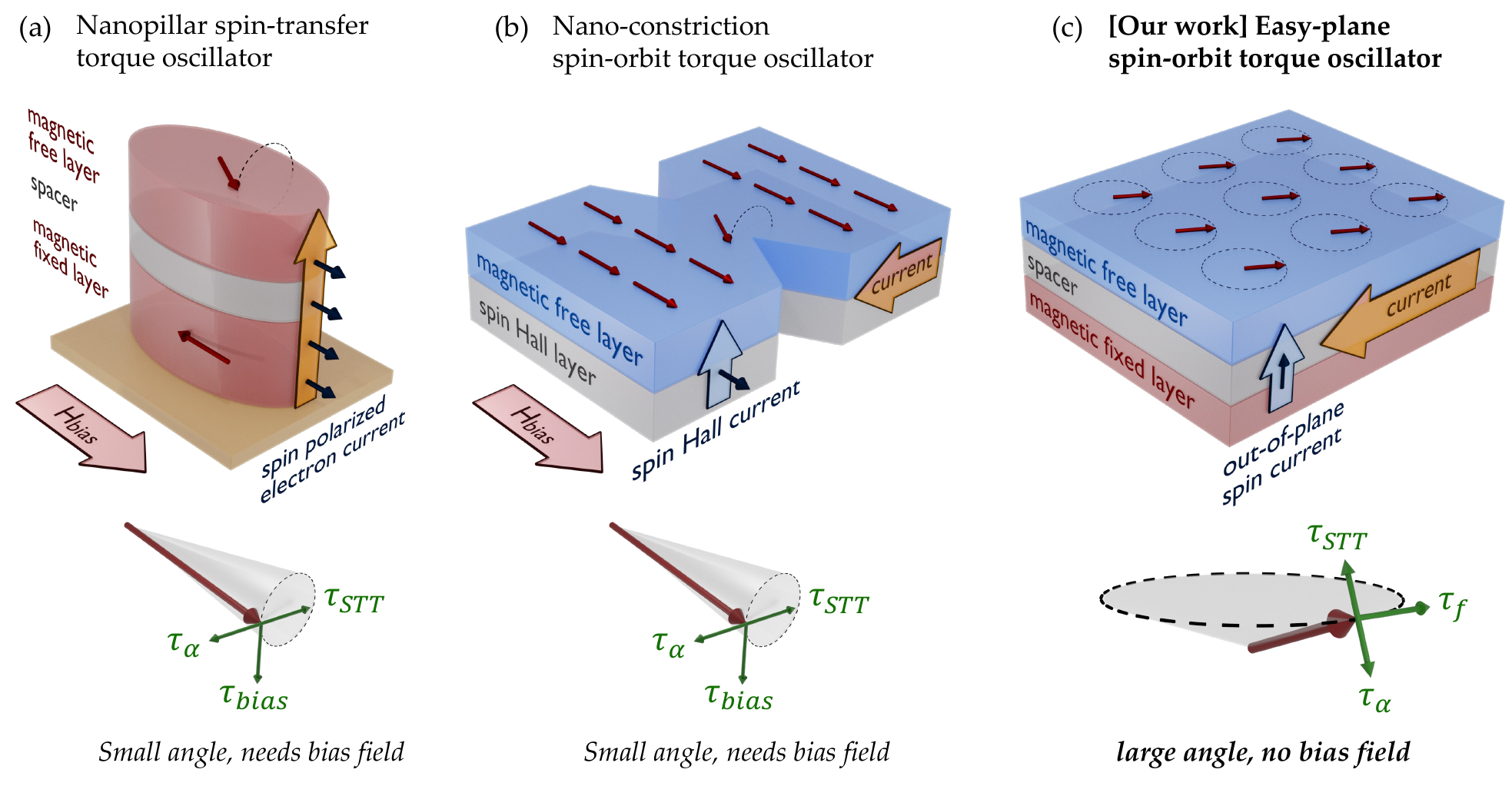} 
   \caption{(a) Conventional spin-transfer torque oscillator with out-of-plane charge current and bias magnetic field. The magnetic fixed layer spin-polarizes the electrical current which is absorbed by the magnetic free layer, imparting a spin-transfer torque. The role of the spin transfer torque is to cancel the damping torque, enabling the free layer magnetization to precess about the bias magnetic field. (b) Conventional spin-orbit torque oscillator with in-plane charge current and bias magnetic field. The bottom layer, typically a heavy metal, converts an in-plane charge current into a spin current flowing out-of-plane. Like (a), the spin current is then absorbed by the magnetic free layer, imparts a spin torque that compensates the damping, and enables magnetization precession about the bias magnetic field. Both (a) and (b) suffer from small output signal, thermal instability, and the requirement of a bias field for operation. (c) Our proposed device, an easy-plane spin-orbit torque oscillator, with in-plane charge current and no bias magnetic field. The in-plane charge current in the magnetic fixed layer generates an out-of-plane spin current that flows to the magnetic free layer and imparts a spin torque. The spin torque tilts the magnetic moments out-of-plane and compensates the material's damping, enabling oscillations around an internal field (e.g., demagnetization field). The torque diagram below each device shows the torques necessary for each to undergo self-sustained oscillations and highlights the large precessional cone angle for the newly proposed device (c).} 
   \label{fig:Devices}
\end{figure*}

To date, there are two major types of spin-torque oscillators. The first is spin-transfer torque oscillators, illustrated in Fig.~\ref{fig:Devices}(a), that are based on nanopillar or nanocontact magnetic tunnel junctions. In this device scheme, a charge current is passed along the vertical axis of the magnetic tunnel junction, which includes a magnetic ``fixed layer.'' The current becomes spin-polarized by the fixed layer and then imparts a spin-transfer torque on the free layer magnetization. A major drawback is that the electric current must pass through a resistive tunnel barrier, which leads to a high power dissipation and durability issues from dielectric breakdown \cite{Yoshida_1}. Another practical disadvantage is that the effective free-layer area must be $\lesssim$ \SI{0.01}{\micro\metre}$^2$ to prevent magnetization curling (e.g., from the current-induced Oersted field) and achieve uniform dynamics for GHz-range output \cite{T_Chen_1,Dumas_1}. The small active area makes the oscillations vulnerable to thermal fluctuations~\cite{Rippard2006}, reducing the oscillator's signal output and quality factor.  

The second type is the spin-orbit torque oscillator, illustrated in Fig.~\ref{fig:Devices}(b), which possesses a simpler planar device structure that overcomes some of the disadvantages of the spin-transfer torque oscillator. A typical spin-orbit torque oscillator consists of a magnetic free layer interfaced with a metal with a strong spin-orbit effect (e.g., spin-Hall effect), such as Pt~\cite{Hoffmann_3, Sinova_SHE_review}. An in-plane charge current generates a spin current, which flows out-of-plane and exerts a ``spin-orbit torque'' on the magnetization in the adjacent free layer~\cite{Manchon_SOT_review}. The driving charge current does not need to pass through a resistive tunnel barrier, thereby permitting lower power consumption and higher device durability~\cite{T_Chen_1, Luqiao_Osc}. Further, the spin-orbit torque oscillator requires a minimum of just two steps of lithography and is easier to fabricate than nanopillars. 
Therefore, substantial effort has been devoted recently to the development of spin-orbit torque oscillators for microwave electronics and neuromorphic computing~\cite{Awad_Synch, Zahedinejad_Synch, Markovic2022, Manna2023, Sethi2023}. 

However, existing spin-orbit torque oscillators exhibit serious drawbacks. First, they have small precessional cone angles of $<$20$^\circ$~\cite{Roadmap2014} and rely on small anisotropic magnetoresistance ratio of $<$1\%~\cite{T_Chen_1, Haidar2021}. Hence, a single spin-orbit torque oscillator typically has a small power output. 
The spin-orbit torque from spin-Hall currents also cannot sustain coherent oscillations over a large lateral area, due to scattering of the coherent mode into different magnon modes~\cite{Duan2014, Demidov2011, Divinskiy2019}. Uniform, coherent oscillations can be stabilized only within a small area of $<$$\SI{0.1}{\micro\metre}^2$, requiring deep-submicron lithography for nanoscale confinement~\cite{Demidov_19, Duan2014} -- e.g., the nano-constriction geometry illustrated in Fig.~\ref{fig:Devices}(b). The small active area results in greater instability from thermal fluctuations~\cite{T_Chen_1}, curtailing the quality-factor~\cite{Demidov_19, Luqiao_Osc, Awad_Synch} or limiting self-oscillations to cryogenic temperatures~\cite{Duan2014}. Lastly, conventional spin-orbit torque oscillators require a bias field to set the axis of precession. While the oscillators themselves could be compact, the need for a bias magnet would make the overall device architecture quite cumbersome.

In this paper, we present micromagnetic simulations on novel \emph{large-amplitude, easy-plane} spin-orbit torque oscillators that overcome the difficulties plaguing the traditional spin-torque oscillators.  
The proposed device resembles a current-in-plane spin valve exhibiting giant magnetoresistance, illustrated in Fig.~\ref{fig:Devices}(c), consisting of a fixed magnetic layer and a free magnetic layer. 
An in-plane direct current in the fixed layer drives coherent easy-plane oscillations in the free layer approaching 90$^\circ$ cone angle over a micron-scale lateral area. 
The key to this device scheme is the spin-orbit effects in the fixed layer to generate an \emph{out-of-plane spin current}, i.e. a spin current with out-of-plane flow and spin orientation. This out-of-plane spin current is guaranteed by symmetry \cite{Seemann_1, Davidson_2, Amin_Review} and quantified in theoretical work exploring both interfacial \cite{Amin_Phenomenology, Amin_Formalism, Amin_Interface, Amin_Review} and bulk \cite{Amin_Interface, Mook_1, Kim_1, Salemi_1} origins. The relevant experimental evidence comes primarily from measurements of spin-orbit torques thought to be induced by out-of-plane spin currents \cite{Humphries_21, BaekAmin_1, SOPE_Hibino, SOPE_Wang, Yang2024}. The spin current flows into the free layer and tilts the magnetic order slightly out-of-plane while opposing the intrinsic damping. The magnetic order precesses about a strong internal effective field rather than an external magnetic field~\cite{Smith_7}, hence permitting precession with within the plane of the free layer [Fig.~\ref{fig:Devices}(c)]. This large-cone-angle precession, inspired by recent proposals of superfluid-like magnetization dynamics~\cite{Sonin2010, Takei2014, Skarsvag2015, Iacocca2019, Liu_23, Smith_7, Shadman2023},  is robust against magnon scattering and can remain coherent over a micron-scale lateral area. 
Thus, this novel spin-orbit torque oscillator is expected to attain (i) a large output signal through a large precession cone angle and giant magnetoresistance, (ii) high stability enabled by a large active area of coherent precession, and (iii) zero-bias-field operation with the precession axis defined by an internal effective field.


An important question is whether realistic spin-orbit effects in the fixed ferromagnetic layer can enable the proposed coherent easy-plane precession. For instance, recent studies indicate that in typical ferromagnetic metals, the magnitude of the out-of-plane spin current is $\sim$10\% of the ``in-plane'' spin-Hall current \cite{Amin_Interface, Salemi_1}. Our micromagnetic simulations demonstrate that coherent easy-plane precession can indeed be realized under such conditions. 
Further, micron-scale coherence of the easy-plane precession is maintained even when thermal fluctuations (corresponding to room temperature) are included. These robust features make the easy-plane spin-orbit torque oscillator a good candidate for modern spintronic applications.


\section{Description of Device}

\subsection{Device geometry and material composition}

Our proposed device has a lateral area on the order of \SI{1}{\micro\metre}$^2$, much greater than the nanopillar and nanoconstriction oscillators. 
In this study, we focus on a free layer comprised of a synthetic antiferromagnet~\cite{Volvach2022}, i.e., \emph{two} ferromagnetic layers coupled antiparallely via the Ruderman-Kittel-Kasuya-Yosida (RKKY) interaction \cite{Stiles_4}.
A free layer consisting of one ferromagnetic film cannot stabilize coherent, self-sustained oscillations over a \SI{}{\micro\metre}$^2$ scale area; dipolar fields from the edges tend to break up a uniform precession mode into multiple precession modes with various phases~\cite{Smith_7, Skarsvag2015}, analogous to the breakup of a single domain magnetization into multiple domains in a large area. The synthetic antiferromagnet greatly reduces the edge dipolar fields via flux closure~\cite{Lepadatu2017}, hence permitting uniform, coherent large cone-angle precession over the large lateral area\footnote{A free layer of synthetic \emph{ferri}magnet, consisting of two ferromagnetic layers with slightly different thicknesses or saturation magnetizations, would also be sufficient for flux-closing the edge dipole fields~\cite{Lepadatu2017} and hence support coherent easy-plane magnetic precession.}. The ability of the synthetic antiferromagnet to stabilize large cone-angle precession was previously demonstrated in simulations of superfluid-like spin transport~\cite{Smith_7, Skarsvag2015}. 

In a macrospin picture, there are three key torques on the magnetization $\mathbf{m}$ in each layer of the synthetic antiferromagnet:
\begin{enumerate}
    \item The spin torque $\mathbf{\tau_{ST}} \propto \mathbf{m} \times (\mathbf{m} \times \mathbf{s})$, from the injected spins $\mathbf{s}$, pulls $\mathbf{m}$ towards  $\mathbf{s}$. Out-of-plane polarized spins $\mathbf{s}||\mathbf{\hat{z}}$ cants the magnetization out-of-plane\footnote{In our micromagnetic simulations, we assume that the spin torque is operative only in the bottom layer (i.e., closer to the fixed layer) of the synthetic antiferromagnet~\cite{Volvach2022}. This is reasonable considering the $\sim$1-nm dephasing length scale of the injected transverse spin current~\cite{Ghosh2012, Lim2022}.}, generating a nonzero $\mathbf{z}$-component of $\mathbf{m}$.  
    \item The field torque $\mathbf{\tau_f} \propto -\mathbf{m} \times \mathbf{B_{eff}}$ causes $\mathbf{m}$ to precess about the net effective field $\mathbf{B_{eff}}$. Here, with the magnetization canted out-of-plane, $\mathbf{B_{eff}}$ consists of the out-of-plane demagnetization field. In the synthetic antiferromagnet, the canted magnetization (misaligned with the other layer's magnetization) experiences an interlayer antiferromagnetic exchange field, which also contributes to $\mathbf{B_{eff}}$. The magnetization sweeps a precessional orbit within the film plane. 
    \item The Gilbert damping torque $\mathbf{\tau_{\alpha}} \sim \mathbf{m} \times  (\mathbf{m} \times \mathbf{B_{eff}})$ pulls $\mathbf{m}$ toward the film plane. In other words, the spin torque $\mathbf{\tau_{ST}}$ must compete with the damping torque $\mathbf{\tau_{\alpha}}$ to cant the magnetization out-of-plane. 
\end{enumerate}

The tilt angle can be increased by increasing the out-of-plane spin current, which is done by increasing the in-plane charge current that generates it. Once the intrinsic damping torque is compensated by the spin torque, the magnetization is free to precess about the internal effective field \cite{Smith_7}. The out-of-plane spin current and the internal field are the key enablers for the proposed device, as they remove the requirement for a bias magnetic field. In this work, we perform full micromagnetic simulations over a finite-sized device that include edge effects and the generation of magnetization textures.

\subsection{Types of spin current injection}

In the magnetic fixed layer where the magnetization is parallel to the applied in-plane electric field (charge current), there are two types of spin currents allowed by symmetry: the spin-Hall current and an out-of-plane spin current. The spin-Hall effect produces a spin current such that the spin flow direction, spin orientation direction, and electric field direction are mutually orthogonal \cite{Hirsch_22}. Both theoretical \cite{Amin_Intrinsic, Miura_1, Zheng_1} and experimental \cite{FMSHEExp_Tian, FMSHEExp_Das, FMSHEExp_Wang, FMSHEExp_Soya} studies indicate that the spin-Hall conductivities of heavy metals (e.g. Pt) and transition metal ferromagnets (e.g. Fe, Co, Ni) and their alloys have similar orders of magnitude. Thus, we expect an appreciable spin-Hall current generated by the fixed magnetic layer.

Due to the lower symmetry of ferromagnets as compared to normal metals, ferromagnets may generate spin currents with less constrained spin orientations. In particular, when the magnetization and electric field are both parallel and in-plane, symmetry allows out-of-plane spin currents to be generated \cite{Seemann_1, Humphries_21, BaekAmin_1}. Such spin currents can arise from multiple microscopic mechanisms, including the spin-orbit precession effect \cite{BaekAmin_1, Amin_Interface, Amin_Review, SOPE_Hibino, SOPE_Wang}, the magnetic spin-Hall effect \cite{Mook_1, Kim_1, Salemi_1}, and spin swapping \cite{Lifshits_1, Pauyac_1, Amin_Review}. 

\begin{figure*}[t] 
   \centering
   \includegraphics[width=0.95\textwidth]{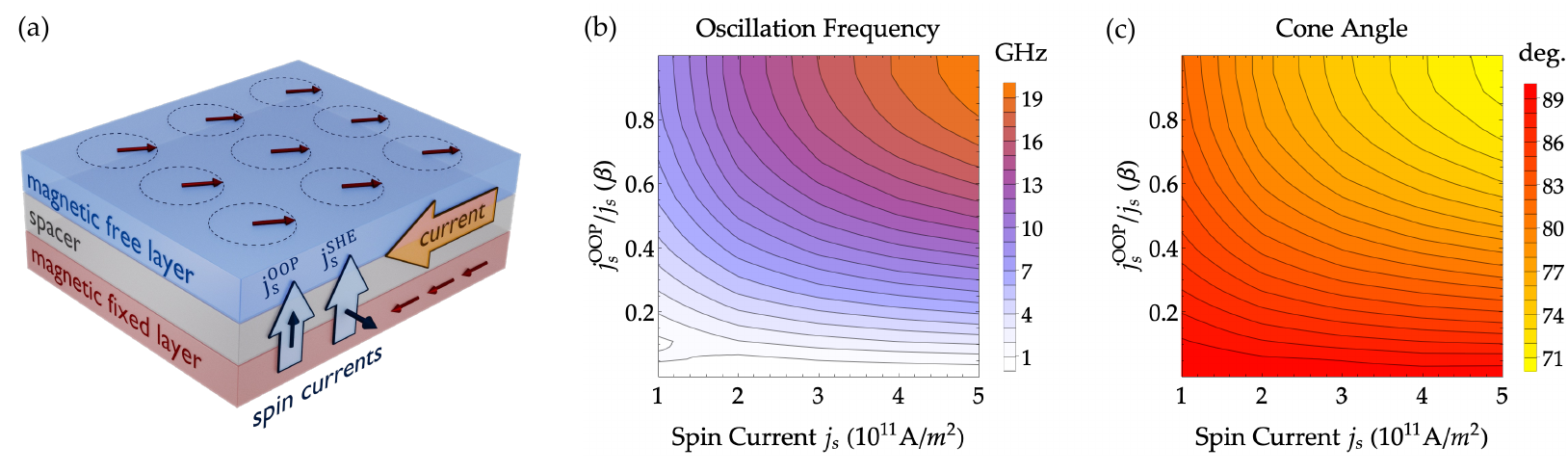} 
   \caption{(a) The proposed device, where the magnetic fixed layer is a ferromagnet and the magnetic free layer is a synthetic antiferromagnet. Red arrows in the fixed layer depict the fixed magnetization while red arrows in the free layer depict the bottom layer magnetization of the synthetic antiferromagnet. Under an applied, in-plane electric field, the fixed layer generates both a spin-Hall current density $\jSHE$ and an out-of-plane spin current density $\jOOP$. (b) Contour plot of the oscillation frequency of the magnetic free layer as a function or the total spin current density $\js$ and the spin current ratio $\ratio$, defined in \Eqs{js}{ratio}. (c) Contour plot of the cone angle of oscillation as a function of the same parameters as in panel (b). Self-sustained oscillations occur over the majority of the parameter space, with the oscillator failing for $\ratio \lesssim 0.1$.}
   \label{fig:ContourPlots300K}
\end{figure*}

We assume that the fixed magnetic layer is the sole source of spin currents\footnote{Note that both the spin-Hall current and out-of-plane spin current can exert self-torques on the ferromagnetic layer that generates them \cite{FMSHEExp_Wang, FMSHEExpTheory_Park, Hoffmann_3}. However, here we are interested in in the case where an out-of-plane spin current escapes the fixed layer and is absorbed in the free layer.}. 
Both the spin-Hall current and the out-of-plane spin current are assumed to be present in the simulated device, as shown in \Fig{fig:ContourPlots300K}(a). These two spin currents have the same out-of-plane flow direction but their spin orientations are orthogonal to each other, leading to a competition of applied torques on the free magnetic layer. The out-of-plane spin current tilts the free layer magnetizations out-of-plane and drives precession, while the spin-Hall current pulls the magnetization in-plane and can perturb the oscillations. The relative strengths of these two torques determine the proposed device's capability to exhibit coherent easy-plane precession.

\begin{figure}[t] 
   \centering
   \includegraphics[width=1.0\columnwidth]{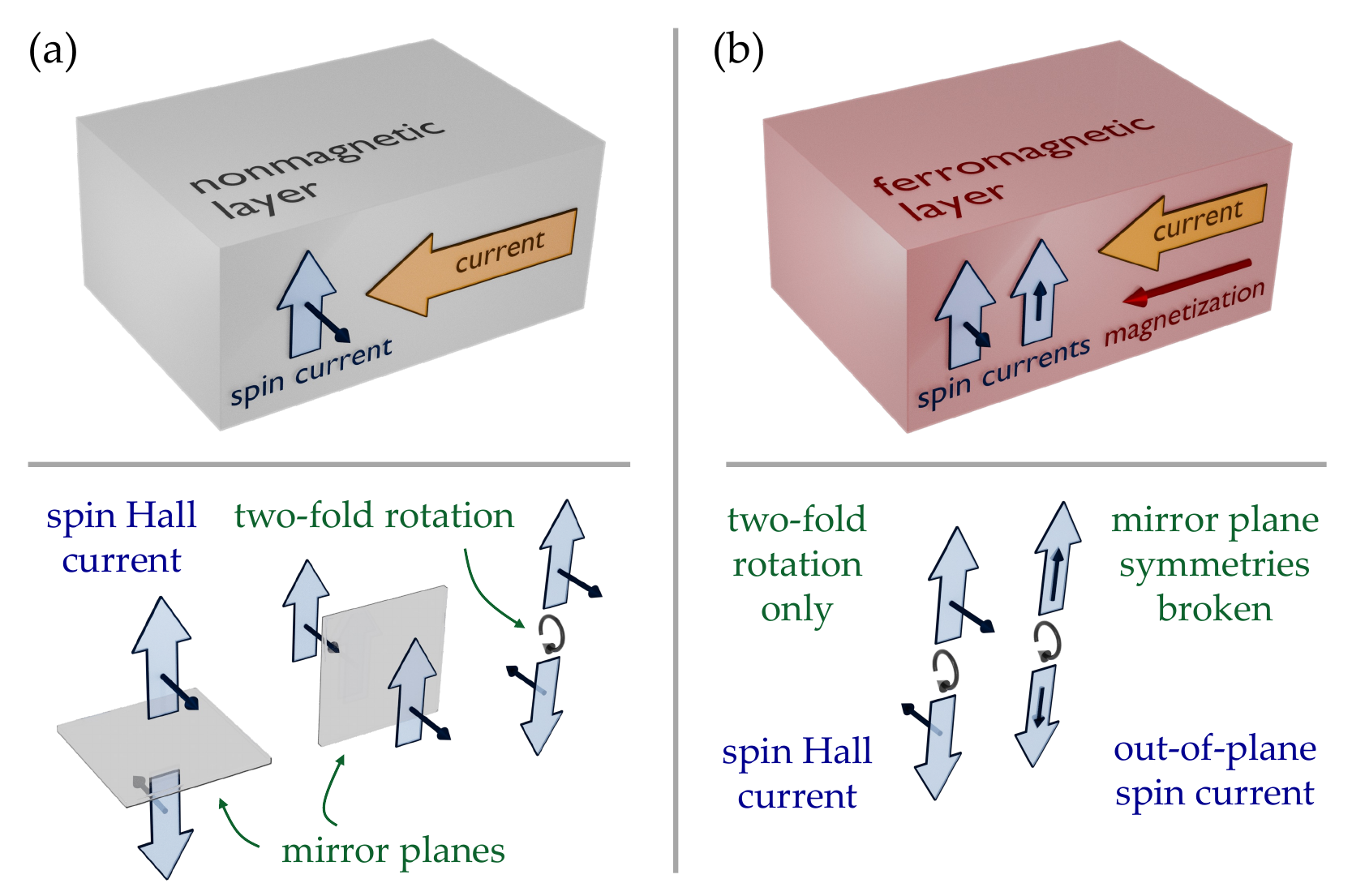} 
   \caption{Spin currents in nonmagnets and ferromagnets allowed by symmetry. a) In bulk nonmagnetic materials under an applied electric field, spin currents satisfy the constraint that the flow direction, spin direction, and electric field are mutually orthogonal. These spin currents, which arise from the spin-Hall effect, are constrained because only this spin current orientation satisfies the crystal's mirror plane and rotational symmetries. b) In bulk ferromagnetic materials, where the applied electric field and magnetization are parallel, the mirror plane symmetries are broken by the magnetization, lowering the symmetry and the constraints on spin currents. Thus, an additional spin current orientation is allowed, where the flow and spin directions are parallel to each other and orthogonal to the electric field and magnetization. In this paper, we focus on such spin currents within a magnetic heterostructure with out-of-plane flow and spin direction, called \emph{out-of-plane spin currents} for short.}
   \label{fig:SymmMech}
\end{figure}

Understanding the microscopic mechanisms responsible for out-of-plane spin currents is not within the scope of this work. While several theoretical predictions suggest that out-of-plane spin current conductivities are comparable to spin-Hall conductivities within ferromagnets, more work is required to confirm these predictions in experiments. Thus, given the uncertainty in the typical strength of out-of-plane spin current generation in ferromagnets and at ferromagnet/nonmagnet interfaces, we simulate various possibilities in this work, from entirely spin-Hall current injection to entirely out-of-plane spin current injection.

\section{Simulation Description}
\label{sec:description}

Our simulations were performed using MuMax$^3$ \cite{MuMax_8}, which calculates the time evolution of a magnetization texture by solving the Landau-Lifshitz-Gilbert (LLG) equation. The LLG equation is given by \cite{LLG_9,Gilbert_10}
\begin{equation}
    \frac{d\mhat}{dt} = \frac{-|\gamma|}{1+\alpha^{2}} \big{(} \mhat\times\bvec{B}_{\text{eff}} + \alpha \mhat\times(\mhat\times\bvec{B}_{\text{eff}}) \big{)} + \torque_S,
\end{equation}
where $\gamma$ is the gyromagnetic ratio, $\alpha$ is the Gilbert damping parameter, $\mhat$ is the magnetization direction, and $\bvec{B_{\text{eff}}}$ is the effective magnetic field. To capture spin torques, we also include the term $\torque_S$, given by Slonczewski \cite{SLONCZEWSKI_5,Finocchio_11},
\begin{equation}
    \torque_S = \frac{g \mu_{B} J g(\theta)}{e M_{\text{sat}}(1+\alpha^2)d} \big{(} \alpha \mhat \times \phat - \frac{1}{M_{\text{sat}}}\mhat\times(\mhat \times \phat) \big{)},
\end{equation}
where $g$ is the Land\'{e} factor, $\mu_B$ is the Bohr magneton, $J$ the charge current density, $\phat$ is the polarization direction of the injected spin current, and $g(\theta)$ \cite{SLONCZEWSKI_5} is given by
\begin{equation}
    g(\theta) = -1 + (1+P)^3 \Big{(} \frac{3+\cos{\theta}}{4P^{3/2}} \Big{)}^{-1},
\end{equation}
where $\theta$ is the angle between $\mhat$ and $\phat$ and $P$ is the polarization of the injected spin current\footnote{Here, $P$ is equivalent to the spin-Hall ratio, i.e., the conversion efficiency of charge current to spin current. For simplicity, we set $P = 1$, but we later comment on the consequence of a more reasonable value of $P$, e.g., of order 0.1.}. 

Figure~\ref{fig:ContourPlots300K}(a) shows the relevant device geometry, where the magnetic fixed layer is a ferromagnet and the magnetic free ``layer'' is a synthetic antiferromagnet. The red arrows in the free layer represent the magnetization of the bottom layer of the synthetic antiferromagnet. 
Each ferromagnetic layer comprising the synthetic antiferromagnet has dimensions \SI{1} {\micro\metre} $\times$ \SI{1} {\micro\metre} $\times$ 2 nm, saturation magnetization 1000 kA/m, Gilbert damping parameter 0.01, and ferromagnetic exchange constant 20~pJ/m. The in-plane magnetocrystalline anisotropy is zero, a good approximation for practical sputter-grown polycrystalline magnetic films. The RKKY interlayer coupling strength between the two layers is -1 mJ/m$^2$. The injected spin current is simulated by including the Slonczewski term $\torque_S$ in the bottom ferromagnetic layer only~\cite{Volvach2022}. 
To make the simulations less cumbersome, we do not explicitly include the fixed layer where the spin current is generated.

The swept parameters are the total spin current density ($\js$) and the ratio ($\ratio$) of out-of-plane spin current density $\jOOP$ to the total spin current density given by 
\begin{align}
    \label{js} \js &= \sqrt{(\jSHE)^2+(\jOOP)^2}    \\
    \label{ratio} \ratio &= \jOOP/\js.
\end{align}
Assuming the electric field (charge current) points along $\xhat$, the spin-Hall current $\jSHE$ has an in-plane spin polarization along $\yhat$. The out-of-plane spin current $\jOOP$ by definition has a spin polarization along $\zhat$. Thus, the polarization direction $\phat$ of the injected spin current lies within the $yz$-plane, where $\arcsin(\ratio)$ is the angle between $\phat$ and $\yhat$. Thus, $\ratio$ determines the polarization angle and $\js$ determines the magnitude of the injected spin current respectively.

The effects of temperature is included by adding a stochastic thermal field to the effective field ($\bvec{B_{\text{eff}}}$) in the LLG equation.\cite{leliaert_12} The stochastic thermal field is given by \cite{MuMax_8,leliaert_12,Brown_13}
\begin{equation}
    \bvec{B}_{\text{therm}} = \boldsymbol{\eta}_{\text{step}}\sqrt{\frac{2\mu_0\alpha k_{B}T}{B_{\text{sat}}\Delta V\Delta t}},
\end{equation}
where $\alpha$ is the damping parameter, $k_B$ is the Boltzmann constant, $T$ is the temperature, $B_{\text{sat}} = \mu_0 M_s$ is the saturation magnetic field, $\Delta V$ is the cell volume, $\Delta t$ is the simulation time step, and $\boldsymbol{\eta}_{\text{step}}$ is a random vector determined via a standard normal distribution.

\section{Results}

\begin{figure*}[t] 
   \centering
   \includegraphics[width=0.925\textwidth]{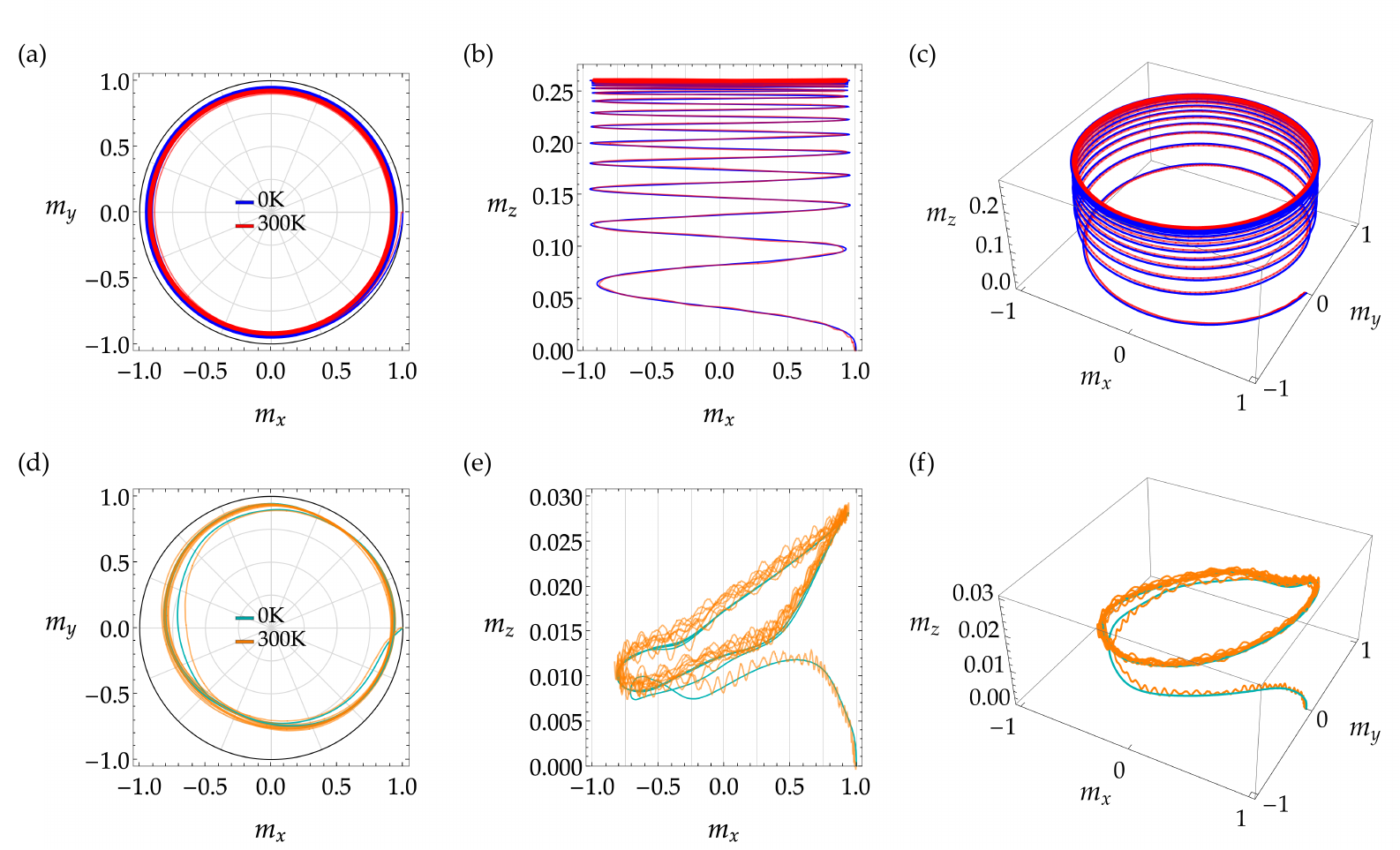} 
   \caption{Trajectories of magnetization direction of one layer in the synthetic antiferromagnet free layer over the unit sphere at absolute zero and at room temperature. The top row panels correspond to $\beta = 1$, that is, the injected spin current from the fixed layer is entirely the out-of-plane spin current. The bottom row panels correspond to $\beta = 0.08$, where the injected spin currents consist of mostly the spin-Hall current. Panels (a) and (d) show the time evolution of the in-plane (i.e. $x$ and $y$) magnetization components, while panels (b) and (e) show the $x$ and $z$ components. Panels (c) and (f) are three-dimensional plots of the same trajectories as (a,b) and (d,e). Panels (a) and (d) show that regardless of temperature and ratio of out-of-plane spin current to spin-Hall current, self-sustained oscillations occur with large cone angle (i.e. with the magnetization mostly in-plane). Panels (b) and (e) show that the effect of temperature is most prominent for low $\beta$ values, and leads to noise in the $z$ component of the magnetization, which has minimal effect on the primary oscillation output signal ($xy$ component of the magnetization).}
   \label{fig:MagTraj}
\end{figure*}


To confirm GHz steady state oscillations in the free layer, simulations were run to \SI{1}{\micro\second} to capture several hundred periods. 
Figure~\ref{fig:ContourPlots300K}(b) shows the oscillation frequency of magnetic free layer as a function of $\js$ and $\ratio$. We choose spin current densities $\js$ on the order of $10^{11}$ A/m$^2$, which are consistent with other micromagnetic simulations of spin-orbit torque oscillators \cite{currentDesnityRange}. Note that the associated charge current densities could be as high as $\sim$$10^{12}$ A/m$^2$ if the spin-Hall ratio of the fixed magnetic layer is only $\sim$0.1. This charge current density is typical for experimentally demonstrated spin-orbit torque oscillators~\cite{Duan2014, Awad_Synch, Divinskiy2019, Zahedinejad_Synch, Demidov_19}. 

As the spin current ratio $\ratio$ is swept from 0 to 1, the injected spin current changes from entirely spin-Hall current $\ratio = 0$ to entirely out-of-plane spin current $\ratio = 1$. The results shown in \Fig{fig:ContourPlots300K}(b) confirm the trend that increasing $\js$ or $\ratio$ increases the oscillation frequency. This trend can be understood as follows. Increasing $\js$ or $\ratio$ will increase the injected out-of-plane spin current (unless $\ratio = 0$), which further tilts the free-layer magnetization out-of-plane. As the tilt angle increases, so does the torque provided by the internal field, which in turn increases the frequency of oscillation. To achieve the highest oscillation frequency within the range of parameters studied, $\js$ and $\ratio$ should be maximized.

\begin{figure*} 
   \centering
   \includegraphics[width=\textwidth]{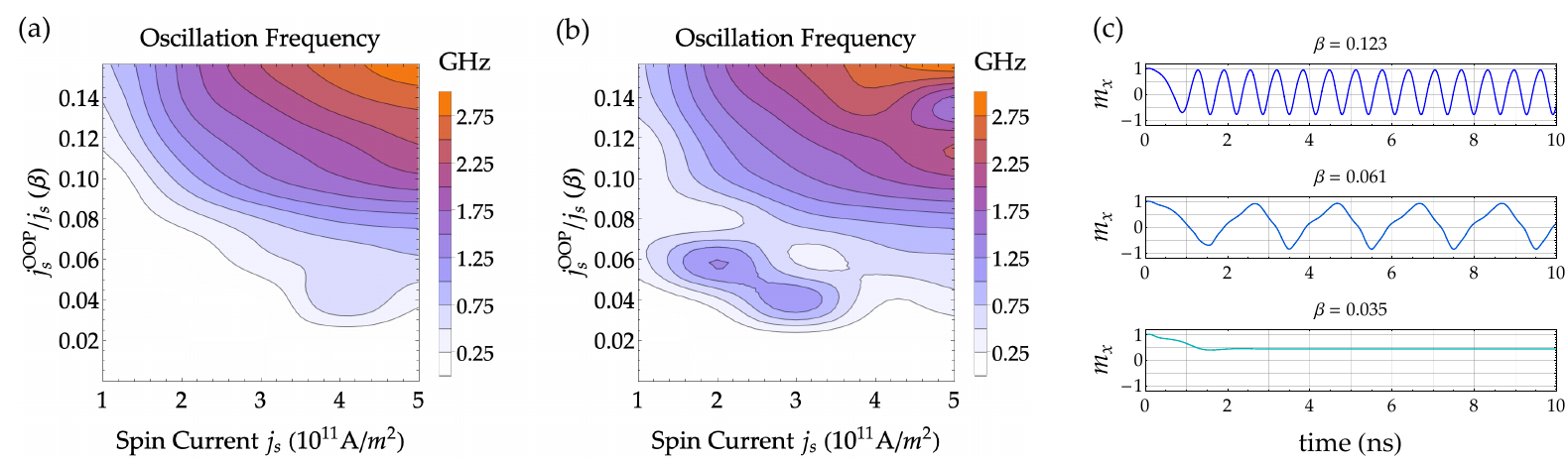} 
   \caption{Threshold regime with low $\beta$ in which self-sustained oscillations emerge. Panel (a) plots oscillation frequency at $T = 0$ K as a function of $\js$ and $\ratio$. Panel (b) is the same as panel (a) but for $T = 300$ K. Panel (c) shows the time evolution of $m_x$ (corresponding to the synthetic antiferromagnet's bottom layer) for $j_s = 3 \times 10^{11} \text{A/m}^2$ for various small $\ratio$ values. Self-sustained oscillations persist even if the out-of-plane spin current is roughly an order of magnitude less than the spin-Hall current.}
   \label{fig:breakdown}
\end{figure*}

Figure~\ref{fig:ContourPlots300K}(c) shows the time-averaged cone angle $\theta_\mathrm{c}$ of magnetic free layer as a function of $\js$ and $\ratio$. Note that a time-averaged cone angle of $\theta_\mathrm{c} = 90^\circ$ corresponds to fully in-plane oscillations. 
The precessing magnetization generates an oscillating voltage output from the swinging resistance, due to giant magnetoresistance of the spin valve. In particular, as the free-layer magnetization (in the bottom layer of the synthetic antiferromagnet, closer to the fixed layer) rotates from being parallel to antiparallel to the fixed-layer magnetization, the resistance swings from its low state to high state~\cite{Coelho2014, Volvach2022}. 
The magnitude of the oscillating signal output is proportional to the in-plane component of the precessing magnetization -- i.e., $\sin \theta_\mathrm{c}$, maximized at $\theta_\mathrm{c} = 90^\circ$. 
By increasing the out-of-plane spin current density, the magnetization tilts further out-of-plane and the time-averaged cone angle decreases. Nevertheless, within our simulated parameter space, the cone angle remains large at $\theta_\mathrm{c} \gtrsim 70^\circ$. The corresponding magnetization remains mostly in-plane ($\sin\theta_\mathrm{c} \gtrsim 0.9$) such that the magnetoresistance signal output remains large. 


In \Fig{fig:MagTraj}, we plot the time evolution of the magnetization direction $\mhat$ at $\js = 3 \times 10^{11} \text{A}/\text{nm}^2$ for both $T = 0$ K and $T = 300$ K and for both $\ratio = 0.08$ (mostly spin-Hall current injection) and $\ratio = 1$ (entirely out-of-plane spin current injection). The magnetization plotted corresponds to the bottom layer of the synthetic antiferromaget. Panels (a) and (d) show that in all cases, the magnetization sweeps a nearly circular path in the $xy$-plane (i.e. in-plane). 

The results shown in \Fig{fig:ContourPlots300K} and \Fig{fig:MagTraj} indicate that large cone-angle, self-sustained GHz oscillations occur in the proposed device over a wide parameter space. 
Panels (b) and (e) in \Fig{fig:MagTraj} highlight the primary effect of temperature, which is to introduce fluctuations in the out-of-plane component of the magnetization, $m_z$. The fluctuations in the in-plane magnetization components, $m_x$ and $m_y$, are only a few percent of the easy-plane precession amplitude. Thus, the thermal fluctuations do not significantly affect the swing in resistance (output voltage) determined by $m_x$ and $m_y$. 

We now discuss the threshold regime in which easy-plane self-oscillations emerge at small $\ratio$ of $\sim$0.1. The out-of-plane spin currents corresponding to such $\ratio$ values are about one order of magnitude less than the spin-Hall currents, well within theoretical predictions \cite{Amin_Interface, Salemi_1}.
To study this threshold regime, we performed additional simulations for small $\beta$ values from $0$ to $0.14$ in increments of $0.02$. As $\beta$ approaches zero, the out-of-plane spin current contribution vanishes, leaving only the spin-Hall current. In this regime, we do not expect easy-plane oscillations to occur, since the out-of-plane spin current is required to tilt the magnetization out-of-plane and induce self-sustained oscillations about the internal field. In \Fig{fig:breakdown}(a)-(b), we show the oscillation frequency as a function of $\beta$ and $j_s$ for (a) 0 K and (b) 300 K in the threshold regime. In both cases, self-sustained oscillations persist for $\beta$ values approaching $0.04$, which suggests that the out-of-plane spin current can be as low as $4\%$ of the spin-Hall current and still create self-sustained oscillations. 

Figure~\ref{fig:breakdown}(c) shows the $x$-component of magnetization plotted versus time to further illustrate dynamics in the threshold regime. While small $\beta$ values can introduce higher-order harmonics, as seen in the plot for $\beta = 0.061$, self-sustained oscillations with large cone angle still persist. At $\beta = 0.123$, higher-order harmonics vanish; such clean sinusoidal oscillations arise from a circular orbit of easy-plane magnetic precession (see Fig.~\ref{fig:ContourPlots300K}(a-c)). Our simulation results indicate that coherent easy-plane precession can be stabilized even at small $\beta$, i.e., when the out-of-plane spin current is only $\sim$10\% of the spin-Hall current. This finding is highly encouraging for realizing large-amplitude, easy-plane spin-orbit torque oscillators under realistic conditions. 

\section{Implications for Applications}
Easy-plane spin-orbit torque oscillators in general exhibit a circular precessional orbit with a large cone angle approaching 90$^\circ$. These features lead to a larger signal output and higher stability against magnon scattering~\cite{Divinskiy2019, Montoya2023} compared to conventional spin-orbit torque oscillators with a small cone-angle, elliptical precessional orbit. Such oscillators are promising for applications in neuromorphic computing~\cite{Markovic2022, Manna2023, Sethi2023} and may be applied to magnetic devices that mimic Josephson junctions~\cite{Takei2017, Liu_23}. 

The existing proposals of easy-plane spin-orbit torque oscillators~\cite{Liu_23,Markovic2022, Manna2023} rely on the anti-damping spin-orbit torque driven by spin-Hall spin current. Hence, the precessional axis is in-plane transverse to the current axis.
To achieve a circular, large-cone-angle precessional orbit, careful tuning of magnetic anisotropy is required. For example, the experimental demonstrations so far~\cite{Divinskiy2019, Montoya2023, Sethi2023} attain easy-plane precession in Co/Ni multilayers with perpendicular magnetic anisotropy, precisely tuned to counterbalance the out-of-plane magnetic shape anisotropy. This approach limits the choices of materials for the free layer, making it difficult to lower damping and enhance magnetoresistance for practical devices. 

In our proposed oscillator driven by out-of-plane spin current, the out-of-plane internal field (e.g., demagnetization field) defines the precessional axis. Hence, the precession is within the film plane -- the natural easy plane for soft ferromagnetic metal films governed by shape anisotropy. No particular engineering of perpendicular magnetic anisotropy is needed, so various materials may be employed to optimize the performance of the oscillator. For instance, Ni-Fe and Fe-V alloys~\cite{Schoen2017_damping, Smith2020, Arora2021, Maizel2024} with low damping and low saturation magnetization may be a good choice to reduce the threshold current density to drive precessional dynamics.  Moreover, our proposed oscillator is essentially based on a current-in-plane spin valve with giant magnetoresistance. Such film heterostructures already find wide usage in commercial sensors and are therefore more amenable to practical applications. The oscillators can then leverage established materials optimization approaches. For example, the giant magnetoresistance ratio may be enhanced to $\sim$10\% -- much greater than the anisotropic magnetoresistance ratio of $\sim$1\% typical for spin-orbit torque oscillators -- with subnanometer interfacial dusting layers~\cite{Parkin1993} and encapsulation with insulating layers~\cite{Swagten1996}. Overall, our proposed scheme is highly promising for broad materials options and compatibility with common device fabrication protocols.  


The biggest practical challenge is to realize a sufficient out-of-plane spin current from the fixed-layer ferromagnet. In a recent experiment, the out-of-spin current was reported to switch perpendicular magnetization in a ``T-type'' current-in-plane spin valve (in-plane fixed layer, out-of-plane free layer)~\cite{BaekAmin_1, Yang2024}. Yet, an experimental report of the out-of-plane spin current tuning or triggering precessional magnetization dynamics is still lacking. 
Symmetry guarantees the emergence of an out-of-plane spin current from an in-plane magnetized ferromagnet~\cite{Seemann_1, Humphries_21, BaekAmin_1}. The question is whether the magnitude of the out-of-plane spin current can become large enough -- particularly under a reasonably low charge current. As discussed in Sec.~\ref{sec:description}, the spin current density $\js$ in our simulations assumes a spin-Hall ratio (conversion efficiency of charge to spin currents) of unity. However, a more reasonable spin-Hall ratio is of order 0.1~\cite{Sinova_SHE_review}, which would yield a charge current density of $\sim$$10^{12}$ A/m$^2$. For real applications. it is desirable to reduce the charge current density by about an order of magnitude to $\sim$$10^{11}$ A/m$^2$. 
Thus, an experimental endeavor should enhance the spin-Hall ratio in ferromagnetic metals for the fixed layer, preferably to $\gtrsim$0.3 reported in some nonmagnetic transition metals~\cite{Pai2012, Demasius2016}. Another approach is to increase the ratio of the out-of-plane to in-plane spin current to $\gg$0.1. These outcomes may be feasible by incorporating elements with strong spin-orbit coupling (e.g., Pt, rare-earth metals) into the fixed-layer ferromagnet~\cite{Hrabec2016}.  

\section{Outlook and Conclusions}
Using micromagnetic simulations, we have demonstrated that self-sustained, large-amplitude GHz oscillations are feasible in spin-orbit torque oscillators without an external bias magnetic field. The spin-orbit torque oscillator consists of a fixed ferromagnetic layer, a spacer layer, and a synthetic antiferromagnet as the magnetic free layer, the latter of which is required to obtain coherent oscillations. The oscillator is driven by an in-plane current, which generates various spin currents in the fixed ferromagnetic layer that flow out-of-plane and exert torques on the magnetic free layer. Oscillations occur about an internal effective field rather than an external magnetic field, with the spin-orbit torque counteracting the damping torque in the free layer. To address the uncertainty in the strength of the relevant spin currents in realistic materials, we varied the ratio of the out-of-plane spin current to the spin-Hall current in our simulations, and found that self-sustained oscillations occur even if the out-of-plane spin current is as low as $\lesssim$10\% of the spin-Hall current. The robust performance of these spin-orbit torque oscillators at room temperature presents intriguing possibilities for future spintronic devices with possible applications to microwave communications and neuromorphic computing.

\begin{acknowledgments}
T.N. was supported by the National Science Foundation under Grant No. DMR-2105219. V.P.A was supported by the National Science Foundation under grant ECCS-2236159. D.A.S. and F.R.-D. were supported by the National Science Foundation under Grant No. DMR-2003914. S.E. was supported by the National Science Foundation under Grant No. ECCS-2236160.
\end{acknowledgments}

\bibliography{bib.bib}

\begin{thebibliography}{84}%
\makeatletter
\providecommand \@ifxundefined [1]{%
 \@ifx{#1\undefined}
}%
\providecommand \@ifnum [1]{%
 \ifnum #1\expandafter \@firstoftwo
 \else \expandafter \@secondoftwo
 \fi
}%
\providecommand \@ifx [1]{%
 \ifx #1\expandafter \@firstoftwo
 \else \expandafter \@secondoftwo
 \fi
}%
\providecommand \natexlab [1]{#1}%
\providecommand \enquote  [1]{``#1''}%
\providecommand \bibnamefont  [1]{#1}%
\providecommand \bibfnamefont [1]{#1}%
\providecommand \citenamefont [1]{#1}%
\providecommand \href@noop [0]{\@secondoftwo}%
\providecommand \href [0]{\begingroup \@sanitize@url \@href}%
\providecommand \@href[1]{\@@startlink{#1}\@@href}%
\providecommand \@@href[1]{\endgroup#1\@@endlink}%
\providecommand \@sanitize@url [0]{\catcode `\\12\catcode `\$12\catcode
  `\&12\catcode `\#12\catcode `\^12\catcode `\_12\catcode `\%12\relax}%
\providecommand \@@startlink[1]{}%
\providecommand \@@endlink[0]{}%
\providecommand \url  [0]{\begingroup\@sanitize@url \@url }%
\providecommand \@url [1]{\endgroup\@href {#1}{\urlprefix }}%
\providecommand \urlprefix  [0]{URL }%
\providecommand \Eprint [0]{\href }%
\providecommand \doibase [0]{https://doi.org/}%
\providecommand \selectlanguage [0]{\@gobble}%
\providecommand \bibinfo  [0]{\@secondoftwo}%
\providecommand \bibfield  [0]{\@secondoftwo}%
\providecommand \translation [1]{[#1]}%
\providecommand \BibitemOpen [0]{}%
\providecommand \bibitemStop [0]{}%
\providecommand \bibitemNoStop [0]{.\EOS\space}%
\providecommand \EOS [0]{\spacefactor3000\relax}%
\providecommand \BibitemShut  [1]{\csname bibitem#1\endcsname}%
\let\auto@bib@innerbib\@empty
\bibitem [{\citenamefont {Locatelli}\ \emph {et~al.}(2014)\citenamefont
  {Locatelli}, \citenamefont {Cros},\ and\ \citenamefont
  {Grollier}}]{Locatelli_15}%
  \BibitemOpen
  \bibfield  {author} {\bibinfo {author} {\bibfnamefont {N.}~\bibnamefont
  {Locatelli}}, \bibinfo {author} {\bibfnamefont {V.}~\bibnamefont {Cros}},\
  and\ \bibinfo {author} {\bibfnamefont {J.}~\bibnamefont {Grollier}},\ }\href
  {https://doi.org/10.1038/nmat3823} {\bibfield  {journal} {\bibinfo  {journal}
  {Nature Materials}\ }\textbf {\bibinfo {volume} {13}},\ \bibinfo {pages} {11}
  (\bibinfo {year} {2014})}\BibitemShut {NoStop}%
\bibitem [{\citenamefont {Grollier}\ \emph {et~al.}(2016)\citenamefont
  {Grollier}, \citenamefont {Querlioz},\ and\ \citenamefont
  {Stiles}}]{Grollier_16}%
  \BibitemOpen
  \bibfield  {author} {\bibinfo {author} {\bibfnamefont {J.}~\bibnamefont
  {Grollier}}, \bibinfo {author} {\bibfnamefont {D.}~\bibnamefont {Querlioz}},\
  and\ \bibinfo {author} {\bibfnamefont {M.~D.}\ \bibnamefont {Stiles}},\
  }\href {https://doi.org/10.1109/JPROC.2016.2597152} {\bibfield  {journal}
  {\bibinfo  {journal} {Proceedings of the IEEE}\ }\textbf {\bibinfo {volume}
  {104}},\ \bibinfo {pages} {2024} (\bibinfo {year} {2016})}\BibitemShut
  {NoStop}%
\bibitem [{\citenamefont {Lim}\ \emph {et~al.}(2013)\citenamefont {Lim},
  \citenamefont {Ahn}, \citenamefont {Kim}, \citenamefont {Lee},\ and\
  \citenamefont {Shin}}]{Lim_20}%
  \BibitemOpen
  \bibfield  {author} {\bibinfo {author} {\bibfnamefont {H.}~\bibnamefont
  {Lim}}, \bibinfo {author} {\bibfnamefont {S.}~\bibnamefont {Ahn}}, \bibinfo
  {author} {\bibfnamefont {M.}~\bibnamefont {Kim}}, \bibinfo {author}
  {\bibfnamefont {S.}~\bibnamefont {Lee}},\ and\ \bibinfo {author}
  {\bibfnamefont {H.}~\bibnamefont {Shin}},\ }\href
  {https://doi.org/10.1155/2013/169312} {\bibfield  {journal} {\bibinfo
  {journal} {Advances in Condensed Matter Physics}\ }\textbf {\bibinfo {volume}
  {2013}} (\bibinfo {year} {2013})}\BibitemShut {NoStop}%
\bibitem [{\citenamefont {Houssameddine}\ \emph {et~al.}(2007)\citenamefont
  {Houssameddine}, \citenamefont {Ebels}, \citenamefont {Dela{\"e}t},
  \citenamefont {Rodmacq}, \citenamefont {Firastrau}, \citenamefont
  {Ponthenier}, \citenamefont {Brunet}, \citenamefont {Thirion}, \citenamefont
  {Michel}, \citenamefont {Prejbeanu-Buda}, \citenamefont {Cyrille},
  \citenamefont {Redon},\ and\ \citenamefont {Dieny}}]{Houssameddine2007}%
  \BibitemOpen
  \bibfield  {author} {\bibinfo {author} {\bibfnamefont {D.}~\bibnamefont
  {Houssameddine}}, \bibinfo {author} {\bibfnamefont {U.}~\bibnamefont
  {Ebels}}, \bibinfo {author} {\bibfnamefont {B.}~\bibnamefont {Dela{\"e}t}},
  \bibinfo {author} {\bibfnamefont {B.}~\bibnamefont {Rodmacq}}, \bibinfo
  {author} {\bibfnamefont {I.}~\bibnamefont {Firastrau}}, \bibinfo {author}
  {\bibfnamefont {F.}~\bibnamefont {Ponthenier}}, \bibinfo {author}
  {\bibfnamefont {M.}~\bibnamefont {Brunet}}, \bibinfo {author} {\bibfnamefont
  {C.}~\bibnamefont {Thirion}}, \bibinfo {author} {\bibfnamefont {J.-P.}\
  \bibnamefont {Michel}}, \bibinfo {author} {\bibfnamefont {L.}~\bibnamefont
  {Prejbeanu-Buda}}, \bibinfo {author} {\bibfnamefont {M.-C.}\ \bibnamefont
  {Cyrille}}, \bibinfo {author} {\bibfnamefont {O.}~\bibnamefont {Redon}},\
  and\ \bibinfo {author} {\bibfnamefont {B.}~\bibnamefont {Dieny}},\ }\href
  {https://doi.org/10.1038/nmat1905} {\bibfield  {journal} {\bibinfo  {journal}
  {Nature Materials}\ }\textbf {\bibinfo {volume} {6}},\ \bibinfo {pages} {447}
  (\bibinfo {year} {2007})}\BibitemShut {NoStop}%
\bibitem [{\citenamefont {Troncoso}\ \emph {et~al.}(2019)\citenamefont
  {Troncoso}, \citenamefont {Rode}, \citenamefont {Stamenov}, \citenamefont
  {Coey},\ and\ \citenamefont {Brataas}}]{Troncoso_17}%
  \BibitemOpen
  \bibfield  {author} {\bibinfo {author} {\bibfnamefont {R.~E.}\ \bibnamefont
  {Troncoso}}, \bibinfo {author} {\bibfnamefont {K.}~\bibnamefont {Rode}},
  \bibinfo {author} {\bibfnamefont {P.}~\bibnamefont {Stamenov}}, \bibinfo
  {author} {\bibfnamefont {J.~M.~D.}\ \bibnamefont {Coey}},\ and\ \bibinfo
  {author} {\bibfnamefont {A.}~\bibnamefont {Brataas}},\ }\href
  {https://doi.org/10.1103/PhysRevB.99.054433} {\bibfield  {journal} {\bibinfo
  {journal} {Phys. Rev. B}\ }\textbf {\bibinfo {volume} {99}},\ \bibinfo
  {pages} {054433} (\bibinfo {year} {2019})}\BibitemShut {NoStop}%
\bibitem [{\citenamefont {Rippard}\ \emph {et~al.}(2004)\citenamefont
  {Rippard}, \citenamefont {Pufall}, \citenamefont {Kaka}, \citenamefont
  {Russek},\ and\ \citenamefont {Silva}}]{Rippard}%
  \BibitemOpen
  \bibfield  {author} {\bibinfo {author} {\bibfnamefont {W.~H.}\ \bibnamefont
  {Rippard}}, \bibinfo {author} {\bibfnamefont {M.~R.}\ \bibnamefont {Pufall}},
  \bibinfo {author} {\bibfnamefont {S.}~\bibnamefont {Kaka}}, \bibinfo {author}
  {\bibfnamefont {S.~E.}\ \bibnamefont {Russek}},\ and\ \bibinfo {author}
  {\bibfnamefont {T.~J.}\ \bibnamefont {Silva}},\ }\href
  {https://doi.org/10.1103/PhysRevLett.92.027201} {\bibfield  {journal}
  {\bibinfo  {journal} {Phys. Rev. Lett.}\ }\textbf {\bibinfo {volume} {92}},\
  \bibinfo {pages} {027201} (\bibinfo {year} {2004})}\BibitemShut {NoStop}%
\bibitem [{\citenamefont {Kiselev}\ \emph {et~al.}(2003)\citenamefont
  {Kiselev}, \citenamefont {Sankey}, \citenamefont {Krivorotov}, \citenamefont
  {Emley}, \citenamefont {Schoelkopf}, \citenamefont {Buhrman},\ and\
  \citenamefont {Ralph}}]{Kiselev2003}%
  \BibitemOpen
  \bibfield  {author} {\bibinfo {author} {\bibfnamefont {S.~I.}\ \bibnamefont
  {Kiselev}}, \bibinfo {author} {\bibfnamefont {J.~C.}\ \bibnamefont {Sankey}},
  \bibinfo {author} {\bibfnamefont {I.~N.}\ \bibnamefont {Krivorotov}},
  \bibinfo {author} {\bibfnamefont {N.~C.}\ \bibnamefont {Emley}}, \bibinfo
  {author} {\bibfnamefont {R.~J.}\ \bibnamefont {Schoelkopf}}, \bibinfo
  {author} {\bibfnamefont {R.~A.}\ \bibnamefont {Buhrman}},\ and\ \bibinfo
  {author} {\bibfnamefont {D.~C.}\ \bibnamefont {Ralph}},\ }\href
  {https://doi.org/10.1038/nature01967} {\bibfield  {journal} {\bibinfo
  {journal} {Nature}\ }\textbf {\bibinfo {volume} {425}},\ \bibinfo {pages}
  {380} (\bibinfo {year} {2003})}\BibitemShut {NoStop}%
\bibitem [{\citenamefont {Slonczewski}(1996)}]{SLONCZEWSKI_5}%
  \BibitemOpen
  \bibfield  {author} {\bibinfo {author} {\bibfnamefont {J.~C.}\ \bibnamefont
  {Slonczewski}},\ }\href
  {https://doi.org/https://doi.org/10.1016/0304-8853(96)00062-5} {\bibfield
  {journal} {\bibinfo  {journal} {Journal of Magnetism and Magnetic Materials}\
  }\textbf {\bibinfo {volume} {159}},\ \bibinfo {pages} {L1} (\bibinfo {year}
  {1996})}\BibitemShut {NoStop}%
\bibitem [{\citenamefont {Stiles}\ and\ \citenamefont
  {Zangwill}(2002)}]{Stiles_6}%
  \BibitemOpen
  \bibfield  {author} {\bibinfo {author} {\bibfnamefont {M.~D.}\ \bibnamefont
  {Stiles}}\ and\ \bibinfo {author} {\bibfnamefont {A.}~\bibnamefont
  {Zangwill}},\ }\href {https://doi.org/10.1103/PhysRevB.66.014407} {\bibfield
  {journal} {\bibinfo  {journal} {Phys. Rev. B}\ }\textbf {\bibinfo {volume}
  {66}},\ \bibinfo {pages} {014407} (\bibinfo {year} {2002})}\BibitemShut
  {NoStop}%
\bibitem [{\citenamefont {Yoshida}\ \emph {et~al.}(2009)\citenamefont
  {Yoshida}, \citenamefont {Kurasawa}, \citenamefont {Lee}, \citenamefont
  {Tsunoda}, \citenamefont {Aoki},\ and\ \citenamefont {Sugiyama}}]{Yoshida_1}%
  \BibitemOpen
  \bibfield  {author} {\bibinfo {author} {\bibfnamefont {C.}~\bibnamefont
  {Yoshida}}, \bibinfo {author} {\bibfnamefont {M.}~\bibnamefont {Kurasawa}},
  \bibinfo {author} {\bibfnamefont {Y.~M.}\ \bibnamefont {Lee}}, \bibinfo
  {author} {\bibfnamefont {K.}~\bibnamefont {Tsunoda}}, \bibinfo {author}
  {\bibfnamefont {M.}~\bibnamefont {Aoki}},\ and\ \bibinfo {author}
  {\bibfnamefont {Y.}~\bibnamefont {Sugiyama}},\ }in\ \href
  {https://doi.org/10.1109/IRPS.2009.5173239} {\emph {\bibinfo {booktitle}
  {2009 IEEE International Reliability Physics Symposium}}}\ (\bibinfo {year}
  {2009})\ pp.\ \bibinfo {pages} {139--142}\BibitemShut {NoStop}%
\bibitem [{\citenamefont {Chen}\ \emph {et~al.}(2016)\citenamefont {Chen},
  \citenamefont {Dumas}, \citenamefont {Eklund}, \citenamefont {Muduli},
  \citenamefont {Houshang}, \citenamefont {Awad}, \citenamefont {Dürrenfeld},
  \citenamefont {Malm}, \citenamefont {Rusu},\ and\ \citenamefont
  {Åkerman}}]{T_Chen_1}%
  \BibitemOpen
  \bibfield  {author} {\bibinfo {author} {\bibfnamefont {T.}~\bibnamefont
  {Chen}}, \bibinfo {author} {\bibfnamefont {R.~K.}\ \bibnamefont {Dumas}},
  \bibinfo {author} {\bibfnamefont {A.}~\bibnamefont {Eklund}}, \bibinfo
  {author} {\bibfnamefont {P.~K.}\ \bibnamefont {Muduli}}, \bibinfo {author}
  {\bibfnamefont {A.}~\bibnamefont {Houshang}}, \bibinfo {author}
  {\bibfnamefont {A.~A.}\ \bibnamefont {Awad}}, \bibinfo {author}
  {\bibfnamefont {P.}~\bibnamefont {Dürrenfeld}}, \bibinfo {author}
  {\bibfnamefont {B.~G.}\ \bibnamefont {Malm}}, \bibinfo {author}
  {\bibfnamefont {A.}~\bibnamefont {Rusu}},\ and\ \bibinfo {author}
  {\bibfnamefont {J.}~\bibnamefont {Åkerman}},\ }\href
  {https://doi.org/10.1109/JPROC.2016.2554518} {\bibfield  {journal} {\bibinfo
  {journal} {Proceedings of the IEEE}\ }\textbf {\bibinfo {volume} {104}},\
  \bibinfo {pages} {1919} (\bibinfo {year} {2016})}\BibitemShut {NoStop}%
\bibitem [{\citenamefont {Dumas}\ \emph {et~al.}(2013)\citenamefont {Dumas},
  \citenamefont {Iacocca}, \citenamefont {Bonetti}, \citenamefont {Sani},
  \citenamefont {Mohseni}, \citenamefont {Eklund}, \citenamefont {Persson},
  \citenamefont {Heinonen},\ and\ \citenamefont {\AA{}kerman}}]{Dumas_1}%
  \BibitemOpen
  \bibfield  {author} {\bibinfo {author} {\bibfnamefont {R.~K.}\ \bibnamefont
  {Dumas}}, \bibinfo {author} {\bibfnamefont {E.}~\bibnamefont {Iacocca}},
  \bibinfo {author} {\bibfnamefont {S.}~\bibnamefont {Bonetti}}, \bibinfo
  {author} {\bibfnamefont {S.~R.}\ \bibnamefont {Sani}}, \bibinfo {author}
  {\bibfnamefont {S.~M.}\ \bibnamefont {Mohseni}}, \bibinfo {author}
  {\bibfnamefont {A.}~\bibnamefont {Eklund}}, \bibinfo {author} {\bibfnamefont
  {J.}~\bibnamefont {Persson}}, \bibinfo {author} {\bibfnamefont
  {O.}~\bibnamefont {Heinonen}},\ and\ \bibinfo {author} {\bibfnamefont
  {J.}~\bibnamefont {\AA{}kerman}},\ }\href
  {https://doi.org/10.1103/PhysRevLett.110.257202} {\bibfield  {journal}
  {\bibinfo  {journal} {Phys. Rev. Lett.}\ }\textbf {\bibinfo {volume} {110}},\
  \bibinfo {pages} {257202} (\bibinfo {year} {2013})}\BibitemShut {NoStop}%
\bibitem [{\citenamefont {Rippard}\ \emph {et~al.}(2006)\citenamefont
  {Rippard}, \citenamefont {Pufall},\ and\ \citenamefont
  {Russek}}]{Rippard2006}%
  \BibitemOpen
  \bibfield  {author} {\bibinfo {author} {\bibfnamefont {W.~H.}\ \bibnamefont
  {Rippard}}, \bibinfo {author} {\bibfnamefont {M.~R.}\ \bibnamefont
  {Pufall}},\ and\ \bibinfo {author} {\bibfnamefont {S.~E.}\ \bibnamefont
  {Russek}},\ }\href {https://doi.org/10.1103/PhysRevB.74.224409} {\bibfield
  {journal} {\bibinfo  {journal} {Physical Review B}\ }\textbf {\bibinfo
  {volume} {74}},\ \bibinfo {pages} {224409} (\bibinfo {year}
  {2006})}\BibitemShut {NoStop}%
\bibitem [{\citenamefont {Hoffmann}(2013)}]{Hoffmann_3}%
  \BibitemOpen
  \bibfield  {author} {\bibinfo {author} {\bibfnamefont {A.}~\bibnamefont
  {Hoffmann}},\ }\href {https://doi.org/10.1109/TMAG.2013.2262947} {\bibfield
  {journal} {\bibinfo  {journal} {IEEE Transactions on Magnetics}\ }\textbf
  {\bibinfo {volume} {49}},\ \bibinfo {pages} {5172} (\bibinfo {year}
  {2013})}\BibitemShut {NoStop}%
\bibitem [{\citenamefont {Sinova}\ \emph {et~al.}(2015)\citenamefont {Sinova},
  \citenamefont {Valenzuela}, \citenamefont {Wunderlich}, \citenamefont
  {Back},\ and\ \citenamefont {Jungwirth}}]{Sinova_SHE_review}%
  \BibitemOpen
  \bibfield  {author} {\bibinfo {author} {\bibfnamefont {J.}~\bibnamefont
  {Sinova}}, \bibinfo {author} {\bibfnamefont {S.~O.}\ \bibnamefont
  {Valenzuela}}, \bibinfo {author} {\bibfnamefont {J.}~\bibnamefont
  {Wunderlich}}, \bibinfo {author} {\bibfnamefont {C.~H.}\ \bibnamefont
  {Back}},\ and\ \bibinfo {author} {\bibfnamefont {T.}~\bibnamefont
  {Jungwirth}},\ }\href {https://doi.org/10.1103/RevModPhys.87.1213} {\bibfield
   {journal} {\bibinfo  {journal} {Reviews of Modern Physics}\ }\textbf
  {\bibinfo {volume} {87}},\ \bibinfo {pages} {1213} (\bibinfo {year}
  {2015})}\BibitemShut {NoStop}%
\bibitem [{\citenamefont {Manchon}\ \emph {et~al.}(2019)\citenamefont
  {Manchon}, \citenamefont {\ifmmode~\check{Z}\else \v{Z}\fi{}elezn\'y},
  \citenamefont {Miron}, \citenamefont {Jungwirth}, \citenamefont {Sinova},
  \citenamefont {Thiaville}, \citenamefont {Garello},\ and\ \citenamefont
  {Gambardella}}]{Manchon_SOT_review}%
  \BibitemOpen
  \bibfield  {author} {\bibinfo {author} {\bibfnamefont {A.}~\bibnamefont
  {Manchon}}, \bibinfo {author} {\bibfnamefont {J.}~\bibnamefont
  {\ifmmode~\check{Z}\else \v{Z}\fi{}elezn\'y}}, \bibinfo {author}
  {\bibfnamefont {I.~M.}\ \bibnamefont {Miron}}, \bibinfo {author}
  {\bibfnamefont {T.}~\bibnamefont {Jungwirth}}, \bibinfo {author}
  {\bibfnamefont {J.}~\bibnamefont {Sinova}}, \bibinfo {author} {\bibfnamefont
  {A.}~\bibnamefont {Thiaville}}, \bibinfo {author} {\bibfnamefont
  {K.}~\bibnamefont {Garello}},\ and\ \bibinfo {author} {\bibfnamefont
  {P.}~\bibnamefont {Gambardella}},\ }\href
  {https://doi.org/10.1103/RevModPhys.91.035004} {\bibfield  {journal}
  {\bibinfo  {journal} {Reviews of Modern Physics}\ }\textbf {\bibinfo {volume}
  {91}},\ \bibinfo {pages} {035004} (\bibinfo {year} {2019})}\BibitemShut
  {NoStop}%
\bibitem [{\citenamefont {Liu}\ \emph {et~al.}(2012)\citenamefont {Liu},
  \citenamefont {Pai}, \citenamefont {Ralph},\ and\ \citenamefont
  {Buhrman}}]{Luqiao_Osc}%
  \BibitemOpen
  \bibfield  {author} {\bibinfo {author} {\bibfnamefont {L.}~\bibnamefont
  {Liu}}, \bibinfo {author} {\bibfnamefont {C.-F.}\ \bibnamefont {Pai}},
  \bibinfo {author} {\bibfnamefont {D.~C.}\ \bibnamefont {Ralph}},\ and\
  \bibinfo {author} {\bibfnamefont {R.~A.}\ \bibnamefont {Buhrman}},\ }\href
  {https://doi.org/10.1103/PhysRevLett.109.186602} {\bibfield  {journal}
  {\bibinfo  {journal} {Physical Review Letters}\ }\textbf {\bibinfo {volume}
  {109}},\ \bibinfo {pages} {186602} (\bibinfo {year} {2012})}\BibitemShut
  {NoStop}%
\bibitem [{\citenamefont {Awad}\ \emph {et~al.}(2017)\citenamefont {Awad},
  \citenamefont {Dürrenfeld}, \citenamefont {Houshang}, \citenamefont
  {Dvornik}, \citenamefont {Iacocca}, \citenamefont {Dumas},\ and\
  \citenamefont {Åkerman}}]{Awad_Synch}%
  \BibitemOpen
  \bibfield  {author} {\bibinfo {author} {\bibfnamefont {A.~A.}\ \bibnamefont
  {Awad}}, \bibinfo {author} {\bibfnamefont {P.}~\bibnamefont {Dürrenfeld}},
  \bibinfo {author} {\bibfnamefont {A.}~\bibnamefont {Houshang}}, \bibinfo
  {author} {\bibfnamefont {M.}~\bibnamefont {Dvornik}}, \bibinfo {author}
  {\bibfnamefont {E.}~\bibnamefont {Iacocca}}, \bibinfo {author} {\bibfnamefont
  {R.~K.}\ \bibnamefont {Dumas}},\ and\ \bibinfo {author} {\bibfnamefont
  {J.}~\bibnamefont {Åkerman}},\ }\href {https://doi.org/10.1038/nphys3927}
  {\bibfield  {journal} {\bibinfo  {journal} {Nature Physics}\ }\textbf
  {\bibinfo {volume} {12}},\ \bibinfo {pages} {292–299} (\bibinfo {year}
  {2017})}\BibitemShut {NoStop}%
\bibitem [{\citenamefont {Zahedinejad}\ \emph {et~al.}(2020)\citenamefont
  {Zahedinejad}, \citenamefont {Awad}, \citenamefont {Muralidhar},
  \citenamefont {Khymyn}, \citenamefont {Fulara}, \citenamefont {Mazraati},
  \citenamefont {Dvornik},\ and\ \citenamefont {Åkerman}}]{Zahedinejad_Synch}%
  \BibitemOpen
  \bibfield  {author} {\bibinfo {author} {\bibfnamefont {M.}~\bibnamefont
  {Zahedinejad}}, \bibinfo {author} {\bibfnamefont {A.~A.}\ \bibnamefont
  {Awad}}, \bibinfo {author} {\bibfnamefont {S.}~\bibnamefont {Muralidhar}},
  \bibinfo {author} {\bibfnamefont {R.}~\bibnamefont {Khymyn}}, \bibinfo
  {author} {\bibfnamefont {H.}~\bibnamefont {Fulara}}, \bibinfo {author}
  {\bibfnamefont {H.}~\bibnamefont {Mazraati}}, \bibinfo {author}
  {\bibfnamefont {M.}~\bibnamefont {Dvornik}},\ and\ \bibinfo {author}
  {\bibfnamefont {J.}~\bibnamefont {Åkerman}},\ }\href
  {https://doi.org/10.1038/s41565-019-0593-9} {\bibfield  {journal} {\bibinfo
  {journal} {Nature Nanotechnology}\ }\textbf {\bibinfo {volume} {15}},\
  \bibinfo {pages} {47–52} (\bibinfo {year} {2020})}\BibitemShut {NoStop}%
\bibitem [{\citenamefont {Marković}\ \emph {et~al.}(2022)\citenamefont
  {Marković}, \citenamefont {Daniels}, \citenamefont {Sethi}, \citenamefont
  {Kent}, \citenamefont {Stiles},\ and\ \citenamefont
  {Grollier}}]{Markovic2022}%
  \BibitemOpen
  \bibfield  {author} {\bibinfo {author} {\bibfnamefont {D.}~\bibnamefont
  {Marković}}, \bibinfo {author} {\bibfnamefont {M.~W.}\ \bibnamefont
  {Daniels}}, \bibinfo {author} {\bibfnamefont {P.}~\bibnamefont {Sethi}},
  \bibinfo {author} {\bibfnamefont {A.~D.}\ \bibnamefont {Kent}}, \bibinfo
  {author} {\bibfnamefont {M.~D.}\ \bibnamefont {Stiles}},\ and\ \bibinfo
  {author} {\bibfnamefont {J.}~\bibnamefont {Grollier}},\ }\href
  {https://doi.org/10.1103/PhysRevB.105.014411} {\bibfield  {journal} {\bibinfo
   {journal} {Phys. Rev. B}\ }\textbf {\bibinfo {volume} {105}},\ \bibinfo
  {pages} {014411} (\bibinfo {year} {2022})}\BibitemShut {NoStop}%
\bibitem [{\citenamefont {Manna}\ \emph {et~al.}(2023)\citenamefont {Manna},
  \citenamefont {Medwal},\ and\ \citenamefont {Rawat}}]{Manna2023}%
  \BibitemOpen
  \bibfield  {author} {\bibinfo {author} {\bibfnamefont {S.}~\bibnamefont
  {Manna}}, \bibinfo {author} {\bibfnamefont {R.}~\bibnamefont {Medwal}},\ and\
  \bibinfo {author} {\bibfnamefont {R.~S.}\ \bibnamefont {Rawat}},\ }\href
  {https://doi.org/10.1103/PhysRevB.108.184411} {\bibfield  {journal} {\bibinfo
   {journal} {Physical Review B}\ }\textbf {\bibinfo {volume} {108}},\ \bibinfo
  {pages} {184411} (\bibinfo {year} {2023})}\BibitemShut {NoStop}%
\bibitem [{\citenamefont {Sethi}\ \emph {et~al.}(2023)\citenamefont {Sethi},
  \citenamefont {Sanz-Hernández}, \citenamefont {Godel}, \citenamefont
  {Krishnia}, \citenamefont {Ajejas}, \citenamefont {Mizrahi}, \citenamefont
  {Cros}, \citenamefont {Marković},\ and\ \citenamefont
  {Grollier}}]{Sethi2023}%
  \BibitemOpen
  \bibfield  {author} {\bibinfo {author} {\bibfnamefont {P.}~\bibnamefont
  {Sethi}}, \bibinfo {author} {\bibfnamefont {D.}~\bibnamefont
  {Sanz-Hernández}}, \bibinfo {author} {\bibfnamefont {F.}~\bibnamefont
  {Godel}}, \bibinfo {author} {\bibfnamefont {S.}~\bibnamefont {Krishnia}},
  \bibinfo {author} {\bibfnamefont {F.}~\bibnamefont {Ajejas}}, \bibinfo
  {author} {\bibfnamefont {A.}~\bibnamefont {Mizrahi}}, \bibinfo {author}
  {\bibfnamefont {V.}~\bibnamefont {Cros}}, \bibinfo {author} {\bibfnamefont
  {D.}~\bibnamefont {Marković}},\ and\ \bibinfo {author} {\bibfnamefont
  {J.}~\bibnamefont {Grollier}},\ }\href
  {https://doi.org/10.1103/PhysRevApplied.19.064018} {\bibfield  {journal}
  {\bibinfo  {journal} {Physical Review Applied}\ }\textbf {\bibinfo {volume}
  {19}},\ \bibinfo {pages} {064018} (\bibinfo {year} {2023})}\BibitemShut
  {NoStop}%
\bibitem [{\citenamefont {Stamps}\ \emph {et~al.}(2014)\citenamefont {Stamps},
  \citenamefont {Breitkreutz}, \citenamefont {Åkerman}, \citenamefont
  {Chumak}, \citenamefont {Otani}, \citenamefont {Bauer}, \citenamefont
  {Thiele}, \citenamefont {Bowen}, \citenamefont {Majetich}, \citenamefont
  {Kläui}, \citenamefont {Prejbeanu}, \citenamefont {Dieny}, \citenamefont
  {Dempsey},\ and\ \citenamefont {Hillebrands}}]{Roadmap2014}%
  \BibitemOpen
  \bibfield  {author} {\bibinfo {author} {\bibfnamefont {R.~L.}\ \bibnamefont
  {Stamps}}, \bibinfo {author} {\bibfnamefont {S.}~\bibnamefont {Breitkreutz}},
  \bibinfo {author} {\bibfnamefont {J.}~\bibnamefont {Åkerman}}, \bibinfo
  {author} {\bibfnamefont {A.~V.}\ \bibnamefont {Chumak}}, \bibinfo {author}
  {\bibfnamefont {Y.}~\bibnamefont {Otani}}, \bibinfo {author} {\bibfnamefont
  {G.~E.~W.}\ \bibnamefont {Bauer}}, \bibinfo {author} {\bibfnamefont {J.-U.}\
  \bibnamefont {Thiele}}, \bibinfo {author} {\bibfnamefont {M.}~\bibnamefont
  {Bowen}}, \bibinfo {author} {\bibfnamefont {S.~A.}\ \bibnamefont {Majetich}},
  \bibinfo {author} {\bibfnamefont {M.}~\bibnamefont {Kläui}}, \bibinfo
  {author} {\bibfnamefont {I.~L.}\ \bibnamefont {Prejbeanu}}, \bibinfo {author}
  {\bibfnamefont {B.}~\bibnamefont {Dieny}}, \bibinfo {author} {\bibfnamefont
  {N.~M.}\ \bibnamefont {Dempsey}},\ and\ \bibinfo {author} {\bibfnamefont
  {B.}~\bibnamefont {Hillebrands}},\ }\href
  {https://doi.org/10.1088/0022-3727/47/33/333001} {\bibfield  {journal}
  {\bibinfo  {journal} {Journal of Physics D: Applied Physics}\ }\textbf
  {\bibinfo {volume} {47}},\ \bibinfo {pages} {333001} (\bibinfo {year}
  {2014})}\BibitemShut {NoStop}%
\bibitem [{\citenamefont {Haidar}\ \emph {et~al.}(2021)\citenamefont {Haidar},
  \citenamefont {Mazraati}, \citenamefont {Dürrenfeld}, \citenamefont
  {Fulara}, \citenamefont {Ranjbar},\ and\ \citenamefont
  {Åkerman}}]{Haidar2021}%
  \BibitemOpen
  \bibfield  {author} {\bibinfo {author} {\bibfnamefont {M.}~\bibnamefont
  {Haidar}}, \bibinfo {author} {\bibfnamefont {H.}~\bibnamefont {Mazraati}},
  \bibinfo {author} {\bibfnamefont {P.}~\bibnamefont {Dürrenfeld}}, \bibinfo
  {author} {\bibfnamefont {H.}~\bibnamefont {Fulara}}, \bibinfo {author}
  {\bibfnamefont {M.}~\bibnamefont {Ranjbar}},\ and\ \bibinfo {author}
  {\bibfnamefont {J.}~\bibnamefont {Åkerman}},\ }\href
  {https://doi.org/10.1063/5.0036098} {\bibfield  {journal} {\bibinfo
  {journal} {Applied Physics Letters}\ }\textbf {\bibinfo {volume} {118}},\
  \bibinfo {pages} {12406} (\bibinfo {year} {2021})}\BibitemShut {NoStop}%
\bibitem [{\citenamefont {Duan}\ \emph {et~al.}(2014)\citenamefont {Duan},
  \citenamefont {Smith}, \citenamefont {Yang}, \citenamefont {Youngblood},
  \citenamefont {Lindner}, \citenamefont {Demidov}, \citenamefont
  {Demokritov},\ and\ \citenamefont {Krivorotov}}]{Duan2014}%
  \BibitemOpen
  \bibfield  {author} {\bibinfo {author} {\bibfnamefont {Z.}~\bibnamefont
  {Duan}}, \bibinfo {author} {\bibfnamefont {A.}~\bibnamefont {Smith}},
  \bibinfo {author} {\bibfnamefont {L.}~\bibnamefont {Yang}}, \bibinfo {author}
  {\bibfnamefont {B.}~\bibnamefont {Youngblood}}, \bibinfo {author}
  {\bibfnamefont {J.}~\bibnamefont {Lindner}}, \bibinfo {author} {\bibfnamefont
  {V.~E.}\ \bibnamefont {Demidov}}, \bibinfo {author} {\bibfnamefont {S.~O.}\
  \bibnamefont {Demokritov}},\ and\ \bibinfo {author} {\bibfnamefont {I.~N.}\
  \bibnamefont {Krivorotov}},\ }\href {https://doi.org/10.1038/ncomms6616}
  {\bibfield  {journal} {\bibinfo  {journal} {Nature Communications}\ }\textbf
  {\bibinfo {volume} {5}},\ \bibinfo {pages} {5616} (\bibinfo {year}
  {2014})}\BibitemShut {NoStop}%
\bibitem [{\citenamefont {Demidov}\ \emph {et~al.}(2011)\citenamefont
  {Demidov}, \citenamefont {Urazhdin}, \citenamefont {Edwards}, \citenamefont
  {Stiles}, \citenamefont {McMichael},\ and\ \citenamefont
  {Demokritov}}]{Demidov2011}%
  \BibitemOpen
  \bibfield  {author} {\bibinfo {author} {\bibfnamefont {V.~E.}\ \bibnamefont
  {Demidov}}, \bibinfo {author} {\bibfnamefont {S.}~\bibnamefont {Urazhdin}},
  \bibinfo {author} {\bibfnamefont {E.~R.~J.}\ \bibnamefont {Edwards}},
  \bibinfo {author} {\bibfnamefont {M.~D.}\ \bibnamefont {Stiles}}, \bibinfo
  {author} {\bibfnamefont {R.~D.}\ \bibnamefont {McMichael}},\ and\ \bibinfo
  {author} {\bibfnamefont {S.~O.}\ \bibnamefont {Demokritov}},\ }\href
  {https://doi.org/10.1103/PhysRevLett.107.107204} {\bibfield  {journal}
  {\bibinfo  {journal} {Physical Review Letters}\ }\textbf {\bibinfo {volume}
  {107}},\ \bibinfo {pages} {107204} (\bibinfo {year} {2011})}\BibitemShut
  {NoStop}%
\bibitem [{\citenamefont {Divinskiy}\ \emph {et~al.}(2019)\citenamefont
  {Divinskiy}, \citenamefont {Urazhdin}, \citenamefont {Demokritov},\ and\
  \citenamefont {Demidov}}]{Divinskiy2019}%
  \BibitemOpen
  \bibfield  {author} {\bibinfo {author} {\bibfnamefont {B.}~\bibnamefont
  {Divinskiy}}, \bibinfo {author} {\bibfnamefont {S.}~\bibnamefont {Urazhdin}},
  \bibinfo {author} {\bibfnamefont {S.~O.}\ \bibnamefont {Demokritov}},\ and\
  \bibinfo {author} {\bibfnamefont {V.~E.}\ \bibnamefont {Demidov}},\ }\href
  {https://doi.org/10.1038/s41467-019-13182-w} {\bibfield  {journal} {\bibinfo
  {journal} {Nature Communications}\ }\textbf {\bibinfo {volume} {10}},\
  \bibinfo {pages} {5211} (\bibinfo {year} {2019})}\BibitemShut {NoStop}%
\bibitem [{\citenamefont {Demidov}\ \emph {et~al.}(2014)\citenamefont
  {Demidov}, \citenamefont {Urazhdin}, \citenamefont {Zholud}, \citenamefont
  {Sadovnikov},\ and\ \citenamefont {Demokritov}}]{Demidov_19}%
  \BibitemOpen
  \bibfield  {author} {\bibinfo {author} {\bibfnamefont {V.~E.}\ \bibnamefont
  {Demidov}}, \bibinfo {author} {\bibfnamefont {S.}~\bibnamefont {Urazhdin}},
  \bibinfo {author} {\bibfnamefont {A.}~\bibnamefont {Zholud}}, \bibinfo
  {author} {\bibfnamefont {A.~V.}\ \bibnamefont {Sadovnikov}},\ and\ \bibinfo
  {author} {\bibfnamefont {S.~O.}\ \bibnamefont {Demokritov}},\ }\href
  {https://doi.org/10.1063/1.4901027} {\bibfield  {journal} {\bibinfo
  {journal} {Applied Physics Letters}\ }\textbf {\bibinfo {volume} {105}},\
  \bibinfo {pages} {172410} (\bibinfo {year} {2014})}\BibitemShut {NoStop}%
\bibitem [{\citenamefont {Seemann}\ \emph {et~al.}(2015)\citenamefont
  {Seemann}, \citenamefont {K\"odderitzsch}, \citenamefont {Wimmer},\ and\
  \citenamefont {Ebert}}]{Seemann_1}%
  \BibitemOpen
  \bibfield  {author} {\bibinfo {author} {\bibfnamefont {M.}~\bibnamefont
  {Seemann}}, \bibinfo {author} {\bibfnamefont {D.}~\bibnamefont
  {K\"odderitzsch}}, \bibinfo {author} {\bibfnamefont {S.}~\bibnamefont
  {Wimmer}},\ and\ \bibinfo {author} {\bibfnamefont {H.}~\bibnamefont
  {Ebert}},\ }\href {https://doi.org/10.1103/PhysRevB.92.155138} {\bibfield
  {journal} {\bibinfo  {journal} {Phys. Rev. B}\ }\textbf {\bibinfo {volume}
  {92}},\ \bibinfo {pages} {155138} (\bibinfo {year} {2015})}\BibitemShut
  {NoStop}%
\bibitem [{\citenamefont {Davidson}\ \emph {et~al.}(2020)\citenamefont
  {Davidson}, \citenamefont {Amin}, \citenamefont {Aljuaid}, \citenamefont
  {Haney},\ and\ \citenamefont {Fan}}]{Davidson_2}%
  \BibitemOpen
  \bibfield  {author} {\bibinfo {author} {\bibfnamefont {A.}~\bibnamefont
  {Davidson}}, \bibinfo {author} {\bibfnamefont {V.~P.}\ \bibnamefont {Amin}},
  \bibinfo {author} {\bibfnamefont {W.~S.}\ \bibnamefont {Aljuaid}}, \bibinfo
  {author} {\bibfnamefont {P.~M.}\ \bibnamefont {Haney}},\ and\ \bibinfo
  {author} {\bibfnamefont {X.}~\bibnamefont {Fan}},\ }\href
  {https://doi.org/https://doi.org/10.1016/j.physleta.2019.126228} {\bibfield
  {journal} {\bibinfo  {journal} {Physics Letters A}\ }\textbf {\bibinfo
  {volume} {384}},\ \bibinfo {pages} {126228} (\bibinfo {year}
  {2020})}\BibitemShut {NoStop}%
\bibitem [{\citenamefont {Amin}\ \emph {et~al.}(2020)\citenamefont {Amin},
  \citenamefont {Haney},\ and\ \citenamefont {Stiles}}]{Amin_Review}%
  \BibitemOpen
  \bibfield  {author} {\bibinfo {author} {\bibfnamefont {V.~P.}\ \bibnamefont
  {Amin}}, \bibinfo {author} {\bibfnamefont {P.~M.}\ \bibnamefont {Haney}},\
  and\ \bibinfo {author} {\bibfnamefont {M.~D.}\ \bibnamefont {Stiles}},\
  }\href {https://doi.org/10.1063/5.0024019} {\bibfield  {journal} {\bibinfo
  {journal} {Journal of Applied Physics}\ }\textbf {\bibinfo {volume} {128}},\
  \bibinfo {pages} {151101} (\bibinfo {year} {2020})}\BibitemShut {NoStop}%
\bibitem [{\citenamefont {Amin}\ and\ \citenamefont
  {Stiles}(2016{\natexlab{a}})}]{Amin_Phenomenology}%
  \BibitemOpen
  \bibfield  {author} {\bibinfo {author} {\bibfnamefont {V.~P.}\ \bibnamefont
  {Amin}}\ and\ \bibinfo {author} {\bibfnamefont {M.~D.}\ \bibnamefont
  {Stiles}},\ }\href {https://doi.org/10.1103/PhysRevB.94.104420} {\bibfield
  {journal} {\bibinfo  {journal} {Phys. Rev. B}\ }\textbf {\bibinfo {volume}
  {94}},\ \bibinfo {pages} {104420} (\bibinfo {year}
  {2016}{\natexlab{a}})}\BibitemShut {NoStop}%
\bibitem [{\citenamefont {Amin}\ and\ \citenamefont
  {Stiles}(2016{\natexlab{b}})}]{Amin_Formalism}%
  \BibitemOpen
  \bibfield  {author} {\bibinfo {author} {\bibfnamefont {V.~P.}\ \bibnamefont
  {Amin}}\ and\ \bibinfo {author} {\bibfnamefont {M.~D.}\ \bibnamefont
  {Stiles}},\ }\href {https://doi.org/10.1103/PhysRevB.94.104419} {\bibfield
  {journal} {\bibinfo  {journal} {Phys. Rev. B}\ }\textbf {\bibinfo {volume}
  {94}},\ \bibinfo {pages} {104419} (\bibinfo {year}
  {2016}{\natexlab{b}})}\BibitemShut {NoStop}%
\bibitem [{\citenamefont {Amin}\ \emph {et~al.}(2018)\citenamefont {Amin},
  \citenamefont {Zemen},\ and\ \citenamefont {Stiles}}]{Amin_Interface}%
  \BibitemOpen
  \bibfield  {author} {\bibinfo {author} {\bibfnamefont {V.~P.}\ \bibnamefont
  {Amin}}, \bibinfo {author} {\bibfnamefont {J.}~\bibnamefont {Zemen}},\ and\
  \bibinfo {author} {\bibfnamefont {M.~D.}\ \bibnamefont {Stiles}},\ }\href
  {https://doi.org/10.1103/PhysRevLett.121.136805} {\bibfield  {journal}
  {\bibinfo  {journal} {Phys. Rev. Lett.}\ }\textbf {\bibinfo {volume} {121}},\
  \bibinfo {pages} {136805} (\bibinfo {year} {2018})}\BibitemShut {NoStop}%
\bibitem [{\citenamefont {Mook}\ \emph {et~al.}(2020)\citenamefont {Mook},
  \citenamefont {Neumann}, \citenamefont {Johansson}, \citenamefont {Henk},\
  and\ \citenamefont {Mertig}}]{Mook_1}%
  \BibitemOpen
  \bibfield  {author} {\bibinfo {author} {\bibfnamefont {A.}~\bibnamefont
  {Mook}}, \bibinfo {author} {\bibfnamefont {R.~R.}\ \bibnamefont {Neumann}},
  \bibinfo {author} {\bibfnamefont {A.}~\bibnamefont {Johansson}}, \bibinfo
  {author} {\bibfnamefont {J.}~\bibnamefont {Henk}},\ and\ \bibinfo {author}
  {\bibfnamefont {I.}~\bibnamefont {Mertig}},\ }\href
  {https://doi.org/10.1103/PhysRevResearch.2.023065} {\bibfield  {journal}
  {\bibinfo  {journal} {Phys. Rev. Res.}\ }\textbf {\bibinfo {volume} {2}},\
  \bibinfo {pages} {023065} (\bibinfo {year} {2020})}\BibitemShut {NoStop}%
\bibitem [{\citenamefont {Kim}\ and\ \citenamefont {Lee}(2020)}]{Kim_1}%
  \BibitemOpen
  \bibfield  {author} {\bibinfo {author} {\bibfnamefont {K.-W.}\ \bibnamefont
  {Kim}}\ and\ \bibinfo {author} {\bibfnamefont {K.-J.}\ \bibnamefont {Lee}},\
  }\href {https://doi.org/10.1103/PhysRevLett.125.207205} {\bibfield  {journal}
  {\bibinfo  {journal} {Phys. Rev. Lett.}\ }\textbf {\bibinfo {volume} {125}},\
  \bibinfo {pages} {207205} (\bibinfo {year} {2020})}\BibitemShut {NoStop}%
\bibitem [{\citenamefont {Salemi}\ and\ \citenamefont
  {Oppeneer}(2022)}]{Salemi_1}%
  \BibitemOpen
  \bibfield  {author} {\bibinfo {author} {\bibfnamefont {L.}~\bibnamefont
  {Salemi}}\ and\ \bibinfo {author} {\bibfnamefont {P.~M.}\ \bibnamefont
  {Oppeneer}},\ }\href {https://doi.org/10.1103/PhysRevB.106.024410} {\bibfield
   {journal} {\bibinfo  {journal} {Phys. Rev. B}\ }\textbf {\bibinfo {volume}
  {106}},\ \bibinfo {pages} {024410} (\bibinfo {year} {2022})}\BibitemShut
  {NoStop}%
\bibitem [{\citenamefont {Humphries}\ \emph {et~al.}(2017)\citenamefont
  {Humphries}, \citenamefont {Wang}, \citenamefont {Edwards}, \citenamefont
  {Allen}, \citenamefont {Shaw}, \citenamefont {Nembach}, \citenamefont {Xiao},
  \citenamefont {Silva},\ and\ \citenamefont {Fan}}]{Humphries_21}%
  \BibitemOpen
  \bibfield  {author} {\bibinfo {author} {\bibfnamefont {A.~M.}\ \bibnamefont
  {Humphries}}, \bibinfo {author} {\bibfnamefont {T.}~\bibnamefont {Wang}},
  \bibinfo {author} {\bibfnamefont {E.~R.~J.}\ \bibnamefont {Edwards}},
  \bibinfo {author} {\bibfnamefont {S.~R.}\ \bibnamefont {Allen}}, \bibinfo
  {author} {\bibfnamefont {J.~M.}\ \bibnamefont {Shaw}}, \bibinfo {author}
  {\bibfnamefont {H.~T.}\ \bibnamefont {Nembach}}, \bibinfo {author}
  {\bibfnamefont {J.~Q.}\ \bibnamefont {Xiao}}, \bibinfo {author}
  {\bibfnamefont {T.~J.}\ \bibnamefont {Silva}},\ and\ \bibinfo {author}
  {\bibfnamefont {X.}~\bibnamefont {Fan}},\ }\href
  {https://doi.org/10.1038/s41467-017-00967-w} {\bibfield  {journal} {\bibinfo
  {journal} {Nature Communications}\ }\textbf {\bibinfo {volume} {8}},\
  \bibinfo {pages} {911} (\bibinfo {year} {2017})}\BibitemShut {NoStop}%
\bibitem [{\citenamefont {Baek}\ \emph {et~al.}(2018)\citenamefont {Baek},
  \citenamefont {Amin}, \citenamefont {Oh}, \citenamefont {Go}, \citenamefont
  {Lee}, \citenamefont {Lee}, \citenamefont {Kim}, \citenamefont {Stiles},
  \citenamefont {Park},\ and\ \citenamefont {Lee}}]{BaekAmin_1}%
  \BibitemOpen
  \bibfield  {author} {\bibinfo {author} {\bibfnamefont {S.-h.~C.}\
  \bibnamefont {Baek}}, \bibinfo {author} {\bibfnamefont {V.~P.}\ \bibnamefont
  {Amin}}, \bibinfo {author} {\bibfnamefont {Y.-W.}\ \bibnamefont {Oh}},
  \bibinfo {author} {\bibfnamefont {G.}~\bibnamefont {Go}}, \bibinfo {author}
  {\bibfnamefont {S.-J.}\ \bibnamefont {Lee}}, \bibinfo {author} {\bibfnamefont
  {G.-H.}\ \bibnamefont {Lee}}, \bibinfo {author} {\bibfnamefont {K.-J.}\
  \bibnamefont {Kim}}, \bibinfo {author} {\bibfnamefont {M.~D.}\ \bibnamefont
  {Stiles}}, \bibinfo {author} {\bibfnamefont {B.-G.}\ \bibnamefont {Park}},\
  and\ \bibinfo {author} {\bibfnamefont {K.-J.}\ \bibnamefont {Lee}},\ }\href
  {https://doi.org/10.1038/s41563-018-0041-5} {\bibfield  {journal} {\bibinfo
  {journal} {Nature Materials}\ }\textbf {\bibinfo {volume} {17}},\ \bibinfo
  {pages} {509} (\bibinfo {year} {2018})}\BibitemShut {NoStop}%
\bibitem [{\citenamefont {Hibino}\ \emph {et~al.}(2020)\citenamefont {Hibino},
  \citenamefont {Hasegawa}, \citenamefont {Koyama},\ and\ \citenamefont
  {Chiba}}]{SOPE_Hibino}%
  \BibitemOpen
  \bibfield  {author} {\bibinfo {author} {\bibfnamefont {Y.}~\bibnamefont
  {Hibino}}, \bibinfo {author} {\bibfnamefont {K.}~\bibnamefont {Hasegawa}},
  \bibinfo {author} {\bibfnamefont {T.}~\bibnamefont {Koyama}},\ and\ \bibinfo
  {author} {\bibfnamefont {D.}~\bibnamefont {Chiba}},\ }\href
  {https://doi.org/10.1063/5.0002326} {\bibfield  {journal} {\bibinfo
  {journal} {APL Materials}\ }\textbf {\bibinfo {volume} {8}},\ \bibinfo
  {pages} {041110} (\bibinfo {year} {2020})}\BibitemShut {NoStop}%
\bibitem [{\citenamefont {Wang}\ \emph {et~al.}(2022)\citenamefont {Wang},
  \citenamefont {Fu}, \citenamefont {Zhou}, \citenamefont {Chen}, \citenamefont
  {Yang}, \citenamefont {Li}, \citenamefont {Tao}, \citenamefont {Yan},
  \citenamefont {Liang}, \citenamefont {Zhan}, \citenamefont {Du},\ and\
  \citenamefont {Liu}}]{SOPE_Wang}%
  \BibitemOpen
  \bibfield  {author} {\bibinfo {author} {\bibfnamefont {W.}~\bibnamefont
  {Wang}}, \bibinfo {author} {\bibfnamefont {Q.}~\bibnamefont {Fu}}, \bibinfo
  {author} {\bibfnamefont {K.}~\bibnamefont {Zhou}}, \bibinfo {author}
  {\bibfnamefont {L.}~\bibnamefont {Chen}}, \bibinfo {author} {\bibfnamefont
  {L.}~\bibnamefont {Yang}}, \bibinfo {author} {\bibfnamefont {Z.}~\bibnamefont
  {Li}}, \bibinfo {author} {\bibfnamefont {Z.}~\bibnamefont {Tao}}, \bibinfo
  {author} {\bibfnamefont {C.}~\bibnamefont {Yan}}, \bibinfo {author}
  {\bibfnamefont {L.}~\bibnamefont {Liang}}, \bibinfo {author} {\bibfnamefont
  {X.}~\bibnamefont {Zhan}}, \bibinfo {author} {\bibfnamefont {Y.}~\bibnamefont
  {Du}},\ and\ \bibinfo {author} {\bibfnamefont {R.}~\bibnamefont {Liu}},\
  }\href {https://doi.org/10.1103/PhysRevApplied.17.034026} {\bibfield
  {journal} {\bibinfo  {journal} {Phys. Rev. Appl.}\ }\textbf {\bibinfo
  {volume} {17}},\ \bibinfo {pages} {034026} (\bibinfo {year}
  {2022})}\BibitemShut {NoStop}%
\bibitem [{\citenamefont {Yang}\ \emph {et~al.}(2024)\citenamefont {Yang},
  \citenamefont {Sun}, \citenamefont {Zeng}, \citenamefont {Cheng},
  \citenamefont {He}, \citenamefont {Yang}, \citenamefont {Wang}, \citenamefont
  {Yu}, \citenamefont {Niu}, \citenamefont {Ji}, \citenamefont {Chen},
  \citenamefont {Miao}, \citenamefont {Wang},\ and\ \citenamefont
  {Ding}}]{Yang2024}%
  \BibitemOpen
  \bibfield  {author} {\bibinfo {author} {\bibfnamefont {M.}~\bibnamefont
  {Yang}}, \bibinfo {author} {\bibfnamefont {L.}~\bibnamefont {Sun}}, \bibinfo
  {author} {\bibfnamefont {Y.}~\bibnamefont {Zeng}}, \bibinfo {author}
  {\bibfnamefont {J.}~\bibnamefont {Cheng}}, \bibinfo {author} {\bibfnamefont
  {K.}~\bibnamefont {He}}, \bibinfo {author} {\bibfnamefont {X.}~\bibnamefont
  {Yang}}, \bibinfo {author} {\bibfnamefont {Z.}~\bibnamefont {Wang}}, \bibinfo
  {author} {\bibfnamefont {L.}~\bibnamefont {Yu}}, \bibinfo {author}
  {\bibfnamefont {H.}~\bibnamefont {Niu}}, \bibinfo {author} {\bibfnamefont
  {T.}~\bibnamefont {Ji}}, \bibinfo {author} {\bibfnamefont {G.}~\bibnamefont
  {Chen}}, \bibinfo {author} {\bibfnamefont {B.}~\bibnamefont {Miao}}, \bibinfo
  {author} {\bibfnamefont {X.}~\bibnamefont {Wang}},\ and\ \bibinfo {author}
  {\bibfnamefont {H.}~\bibnamefont {Ding}},\ }\href
  {https://doi.org/10.1038/s41467-024-07834-w} {\bibfield  {journal} {\bibinfo
  {journal} {Nat. Commun.}\ }\textbf {\bibinfo {volume} {15}},\ \bibinfo
  {pages} {3201} (\bibinfo {year} {2024})}\BibitemShut {NoStop}%
\bibitem [{\citenamefont {Smith}\ \emph {et~al.}(2021)\citenamefont {Smith},
  \citenamefont {Takei}, \citenamefont {Brann}, \citenamefont {Compton},
  \citenamefont {Ramos-Diaz}, \citenamefont {Simmers},\ and\ \citenamefont
  {Emori}}]{Smith_7}%
  \BibitemOpen
  \bibfield  {author} {\bibinfo {author} {\bibfnamefont {D.~A.}\ \bibnamefont
  {Smith}}, \bibinfo {author} {\bibfnamefont {S.}~\bibnamefont {Takei}},
  \bibinfo {author} {\bibfnamefont {B.}~\bibnamefont {Brann}}, \bibinfo
  {author} {\bibfnamefont {L.}~\bibnamefont {Compton}}, \bibinfo {author}
  {\bibfnamefont {F.}~\bibnamefont {Ramos-Diaz}}, \bibinfo {author}
  {\bibfnamefont {M.~J.}\ \bibnamefont {Simmers}},\ and\ \bibinfo {author}
  {\bibfnamefont {S.}~\bibnamefont {Emori}},\ }\href
  {https://doi.org/10.1103/PhysRevApplied.16.054002} {\bibfield  {journal}
  {\bibinfo  {journal} {Phys. Rev. Applied}\ }\textbf {\bibinfo {volume}
  {16}},\ \bibinfo {pages} {054002} (\bibinfo {year} {2021})}\BibitemShut
  {NoStop}%
\bibitem [{\citenamefont {Sonin}(2010)}]{Sonin2010}%
  \BibitemOpen
  \bibfield  {author} {\bibinfo {author} {\bibfnamefont {E.~B.}\ \bibnamefont
  {Sonin}},\ }\href {https://doi.org/10.1080/00018731003739943} {\bibfield
  {journal} {\bibinfo  {journal} {Advances in Physics}\ }\textbf {\bibinfo
  {volume} {59}},\ \bibinfo {pages} {181} (\bibinfo {year} {2010})}\BibitemShut
  {NoStop}%
\bibitem [{\citenamefont {Takei}\ and\ \citenamefont
  {Tserkovnyak}(2014)}]{Takei2014}%
  \BibitemOpen
  \bibfield  {author} {\bibinfo {author} {\bibfnamefont {S.}~\bibnamefont
  {Takei}}\ and\ \bibinfo {author} {\bibfnamefont {Y.}~\bibnamefont
  {Tserkovnyak}},\ }\href {https://doi.org/10.1103/PhysRevLett.112.227201}
  {\bibfield  {journal} {\bibinfo  {journal} {Physical Review Letters}\
  }\textbf {\bibinfo {volume} {112}},\ \bibinfo {pages} {227201} (\bibinfo
  {year} {2014})}\BibitemShut {NoStop}%
\bibitem [{\citenamefont {Skarsvåg}\ \emph {et~al.}(2015)\citenamefont
  {Skarsvåg}, \citenamefont {Holmqvist},\ and\ \citenamefont
  {Brataas}}]{Skarsvag2015}%
  \BibitemOpen
  \bibfield  {author} {\bibinfo {author} {\bibfnamefont {H.}~\bibnamefont
  {Skarsvåg}}, \bibinfo {author} {\bibfnamefont {C.}~\bibnamefont
  {Holmqvist}},\ and\ \bibinfo {author} {\bibfnamefont {A.}~\bibnamefont
  {Brataas}},\ }\href {https://doi.org/10.1103/PhysRevLett.115.237201}
  {\bibfield  {journal} {\bibinfo  {journal} {Phys. Rev. Lett.}\ }\textbf
  {\bibinfo {volume} {115}},\ \bibinfo {pages} {237201} (\bibinfo {year}
  {2015})}\BibitemShut {NoStop}%
\bibitem [{\citenamefont {Iacocca}\ and\ \citenamefont
  {Hoefer}(2019)}]{Iacocca2019}%
  \BibitemOpen
  \bibfield  {author} {\bibinfo {author} {\bibfnamefont {E.}~\bibnamefont
  {Iacocca}}\ and\ \bibinfo {author} {\bibfnamefont {M.~A.}\ \bibnamefont
  {Hoefer}},\ }\href {https://doi.org/10.1016/j.physleta.2019.125858}
  {\bibfield  {journal} {\bibinfo  {journal} {Physics Letters A}\ ,\ \bibinfo
  {pages} {125858}} (\bibinfo {year} {2019})}\BibitemShut {NoStop}%
\bibitem [{\citenamefont {Liu}\ \emph {et~al.}(2020)\citenamefont {Liu},
  \citenamefont {Barsukov}, \citenamefont {Barlas}, \citenamefont
  {Krivorotov},\ and\ \citenamefont {Lake}}]{Liu_23}%
  \BibitemOpen
  \bibfield  {author} {\bibinfo {author} {\bibfnamefont {Y.}~\bibnamefont
  {Liu}}, \bibinfo {author} {\bibfnamefont {I.}~\bibnamefont {Barsukov}},
  \bibinfo {author} {\bibfnamefont {Y.}~\bibnamefont {Barlas}}, \bibinfo
  {author} {\bibfnamefont {I.~N.}\ \bibnamefont {Krivorotov}},\ and\ \bibinfo
  {author} {\bibfnamefont {R.~K.}\ \bibnamefont {Lake}},\ }\href
  {https://doi.org/10.1063/5.0003477} {\bibfield  {journal} {\bibinfo
  {journal} {Applied Physics Letters}\ }\textbf {\bibinfo {volume} {116}},\
  \bibinfo {pages} {132409} (\bibinfo {year} {2020})}\BibitemShut {NoStop}%
\bibitem [{\citenamefont {Shadman}\ and\ \citenamefont
  {Zhu}(2023)}]{Shadman2023}%
  \BibitemOpen
  \bibfield  {author} {\bibinfo {author} {\bibfnamefont {A.}~\bibnamefont
  {Shadman}}\ and\ \bibinfo {author} {\bibfnamefont {J.-G.}\ \bibnamefont
  {Zhu}},\ }\href {https://doi.org/10.1038/s41598-023-13438-5} {\bibfield
  {journal} {\bibinfo  {journal} {Scientific Reports}\ }\textbf {\bibinfo
  {volume} {13}},\ \bibinfo {pages} {13438} (\bibinfo {year}
  {2023})}\BibitemShut {NoStop}%
\bibitem [{\citenamefont {Volvach}\ \emph {et~al.}(2022)\citenamefont
  {Volvach}, \citenamefont {Kent}, \citenamefont {Fullerton},\ and\
  \citenamefont {Lomakin}}]{Volvach2022}%
  \BibitemOpen
  \bibfield  {author} {\bibinfo {author} {\bibfnamefont {I.}~\bibnamefont
  {Volvach}}, \bibinfo {author} {\bibfnamefont {A.~D.}\ \bibnamefont {Kent}},
  \bibinfo {author} {\bibfnamefont {E.~E.}\ \bibnamefont {Fullerton}},\ and\
  \bibinfo {author} {\bibfnamefont {V.}~\bibnamefont {Lomakin}},\ }\href
  {https://doi.org/10.1103/PhysRevApplied.18.024071} {\bibfield  {journal}
  {\bibinfo  {journal} {Phys. Rev. Applied}\ }\textbf {\bibinfo {volume}
  {18}},\ \bibinfo {pages} {024071} (\bibinfo {year} {2022})}\BibitemShut
  {NoStop}%
\bibitem [{\citenamefont {Duine}\ \emph {et~al.}(2018)\citenamefont {Duine},
  \citenamefont {Lee}, \citenamefont {Parkin},\ and\ \citenamefont
  {Stiles}}]{Stiles_4}%
  \BibitemOpen
  \bibfield  {author} {\bibinfo {author} {\bibfnamefont {R.}~\bibnamefont
  {Duine}}, \bibinfo {author} {\bibfnamefont {K.}~\bibnamefont {Lee}}, \bibinfo
  {author} {\bibfnamefont {S.}~\bibnamefont {Parkin}},\ and\ \bibinfo {author}
  {\bibfnamefont {M.}~\bibnamefont {Stiles}},\ }\href
  {https://doi.org/https://doi.org/10.1038/s41567-018-0050-y} {\bibfield
  {journal} {\bibinfo  {journal} {Nature Physics}\ }\textbf {\bibinfo {volume}
  {14}},\ \bibinfo {pages} {217–219} (\bibinfo {year} {2018})}\BibitemShut
  {NoStop}%
\bibitem [{\citenamefont {Lepadatu}\ \emph {et~al.}(2017)\citenamefont
  {Lepadatu}, \citenamefont {Saarikoski}, \citenamefont {Beacham},
  \citenamefont {Benitez}, \citenamefont {Moore}, \citenamefont {Burnell},
  \citenamefont {Sugimoto}, \citenamefont {Yesudas}, \citenamefont {Wheeler},
  \citenamefont {Miguel}, \citenamefont {Dhesi}, \citenamefont {McGrouther},
  \citenamefont {McVitie}, \citenamefont {Tatara},\ and\ \citenamefont
  {Marrows}}]{Lepadatu2017}%
  \BibitemOpen
  \bibfield  {author} {\bibinfo {author} {\bibfnamefont {S.}~\bibnamefont
  {Lepadatu}}, \bibinfo {author} {\bibfnamefont {H.}~\bibnamefont
  {Saarikoski}}, \bibinfo {author} {\bibfnamefont {R.}~\bibnamefont {Beacham}},
  \bibinfo {author} {\bibfnamefont {M.~J.}\ \bibnamefont {Benitez}}, \bibinfo
  {author} {\bibfnamefont {T.~A.}\ \bibnamefont {Moore}}, \bibinfo {author}
  {\bibfnamefont {G.}~\bibnamefont {Burnell}}, \bibinfo {author} {\bibfnamefont
  {S.}~\bibnamefont {Sugimoto}}, \bibinfo {author} {\bibfnamefont
  {D.}~\bibnamefont {Yesudas}}, \bibinfo {author} {\bibfnamefont {M.~C.}\
  \bibnamefont {Wheeler}}, \bibinfo {author} {\bibfnamefont {J.}~\bibnamefont
  {Miguel}}, \bibinfo {author} {\bibfnamefont {S.~S.}\ \bibnamefont {Dhesi}},
  \bibinfo {author} {\bibfnamefont {D.}~\bibnamefont {McGrouther}}, \bibinfo
  {author} {\bibfnamefont {S.}~\bibnamefont {McVitie}}, \bibinfo {author}
  {\bibfnamefont {G.}~\bibnamefont {Tatara}},\ and\ \bibinfo {author}
  {\bibfnamefont {C.~H.}\ \bibnamefont {Marrows}},\ }\href
  {https://doi.org/10.1038/s41598-017-01868-1} {\bibfield  {journal} {\bibinfo
  {journal} {Sci. Rep.}\ }\textbf {\bibinfo {volume} {7}},\ \bibinfo {pages}
  {1640} (\bibinfo {year} {2017})}\BibitemShut {NoStop}%
\bibitem [{\citenamefont {Ghosh}\ \emph {et~al.}(2012)\citenamefont {Ghosh},
  \citenamefont {Auffret}, \citenamefont {Ebels},\ and\ \citenamefont
  {Bailey}}]{Ghosh2012}%
  \BibitemOpen
  \bibfield  {author} {\bibinfo {author} {\bibfnamefont {A.}~\bibnamefont
  {Ghosh}}, \bibinfo {author} {\bibfnamefont {S.}~\bibnamefont {Auffret}},
  \bibinfo {author} {\bibfnamefont {U.}~\bibnamefont {Ebels}},\ and\ \bibinfo
  {author} {\bibfnamefont {W.~E.}\ \bibnamefont {Bailey}},\ }\href
  {https://doi.org/10.1103/PhysRevLett.109.127202} {\bibfield  {journal}
  {\bibinfo  {journal} {Phys. Rev. Lett.}\ }\textbf {\bibinfo {volume} {109}},\
  \bibinfo {pages} {127202} (\bibinfo {year} {2012})}\BibitemShut {NoStop}%
\bibitem [{\citenamefont {Lim}\ \emph {et~al.}(2022)\citenamefont {Lim},
  \citenamefont {Wu}, \citenamefont {Smith}, \citenamefont {Klewe},
  \citenamefont {Shafer},\ and\ \citenamefont {Emori}}]{Lim2022}%
  \BibitemOpen
  \bibfield  {author} {\bibinfo {author} {\bibfnamefont {Y.}~\bibnamefont
  {Lim}}, \bibinfo {author} {\bibfnamefont {S.}~\bibnamefont {Wu}}, \bibinfo
  {author} {\bibfnamefont {D.~A.}\ \bibnamefont {Smith}}, \bibinfo {author}
  {\bibfnamefont {C.}~\bibnamefont {Klewe}}, \bibinfo {author} {\bibfnamefont
  {P.}~\bibnamefont {Shafer}},\ and\ \bibinfo {author} {\bibfnamefont
  {S.}~\bibnamefont {Emori}},\ }\href {https://doi.org/10.1063/5.0071079}
  {\bibfield  {journal} {\bibinfo  {journal} {Appl. Phys. Lett.}\ }\textbf
  {\bibinfo {volume} {121}},\ \bibinfo {pages} {222403} (\bibinfo {year}
  {2022})}\BibitemShut {NoStop}%
\bibitem [{\citenamefont {Hirsch}(1999)}]{Hirsch_22}%
  \BibitemOpen
  \bibfield  {author} {\bibinfo {author} {\bibfnamefont {J.~E.}\ \bibnamefont
  {Hirsch}},\ }\href {https://doi.org/10.1103/PhysRevLett.83.1834} {\bibfield
  {journal} {\bibinfo  {journal} {Phys. Rev. Lett.}\ }\textbf {\bibinfo
  {volume} {83}},\ \bibinfo {pages} {1834} (\bibinfo {year}
  {1999})}\BibitemShut {NoStop}%
\bibitem [{\citenamefont {Amin}\ \emph {et~al.}(2019)\citenamefont {Amin},
  \citenamefont {Li}, \citenamefont {Stiles},\ and\ \citenamefont
  {Haney}}]{Amin_Intrinsic}%
  \BibitemOpen
  \bibfield  {author} {\bibinfo {author} {\bibfnamefont {V.~P.}\ \bibnamefont
  {Amin}}, \bibinfo {author} {\bibfnamefont {J.}~\bibnamefont {Li}}, \bibinfo
  {author} {\bibfnamefont {M.~D.}\ \bibnamefont {Stiles}},\ and\ \bibinfo
  {author} {\bibfnamefont {P.~M.}\ \bibnamefont {Haney}},\ }\href
  {https://doi.org/10.1103/PhysRevB.99.220405} {\bibfield  {journal} {\bibinfo
  {journal} {Phys. Rev. B}\ }\textbf {\bibinfo {volume} {99}},\ \bibinfo
  {pages} {220405} (\bibinfo {year} {2019})}\BibitemShut {NoStop}%
\bibitem [{\citenamefont {Miura}\ and\ \citenamefont {Masuda}(2021)}]{Miura_1}%
  \BibitemOpen
  \bibfield  {author} {\bibinfo {author} {\bibfnamefont {Y.}~\bibnamefont
  {Miura}}\ and\ \bibinfo {author} {\bibfnamefont {K.}~\bibnamefont {Masuda}},\
  }\href {https://doi.org/10.1103/PhysRevMaterials.5.L101402} {\bibfield
  {journal} {\bibinfo  {journal} {Phys. Rev. Mater.}\ }\textbf {\bibinfo
  {volume} {5}},\ \bibinfo {pages} {L101402} (\bibinfo {year}
  {2021})}\BibitemShut {NoStop}%
\bibitem [{\citenamefont {Zheng}\ \emph {et~al.}(2024)\citenamefont {Zheng},
  \citenamefont {Zhu}, \citenamefont {Dong}, \citenamefont {Li}, \citenamefont
  {Zhou}, \citenamefont {Wu},\ and\ \citenamefont {Zhang}}]{Zheng_1}%
  \BibitemOpen
  \bibfield  {author} {\bibinfo {author} {\bibfnamefont {F.}~\bibnamefont
  {Zheng}}, \bibinfo {author} {\bibfnamefont {M.}~\bibnamefont {Zhu}}, \bibinfo
  {author} {\bibfnamefont {J.}~\bibnamefont {Dong}}, \bibinfo {author}
  {\bibfnamefont {X.}~\bibnamefont {Li}}, \bibinfo {author} {\bibfnamefont
  {Y.}~\bibnamefont {Zhou}}, \bibinfo {author} {\bibfnamefont {K.}~\bibnamefont
  {Wu}},\ and\ \bibinfo {author} {\bibfnamefont {J.}~\bibnamefont {Zhang}},\
  }\href {https://doi.org/10.1103/PhysRevB.109.224401} {\bibfield  {journal}
  {\bibinfo  {journal} {Phys. Rev. B}\ }\textbf {\bibinfo {volume} {109}},\
  \bibinfo {pages} {224401} (\bibinfo {year} {2024})}\BibitemShut {NoStop}%
\bibitem [{\citenamefont {Tian}\ \emph {et~al.}(2016)\citenamefont {Tian},
  \citenamefont {Li}, \citenamefont {Qu}, \citenamefont {Huang}, \citenamefont
  {Jin},\ and\ \citenamefont {Chien}}]{FMSHEExp_Tian}%
  \BibitemOpen
  \bibfield  {author} {\bibinfo {author} {\bibfnamefont {D.}~\bibnamefont
  {Tian}}, \bibinfo {author} {\bibfnamefont {Y.}~\bibnamefont {Li}}, \bibinfo
  {author} {\bibfnamefont {D.}~\bibnamefont {Qu}}, \bibinfo {author}
  {\bibfnamefont {S.~Y.}\ \bibnamefont {Huang}}, \bibinfo {author}
  {\bibfnamefont {X.}~\bibnamefont {Jin}},\ and\ \bibinfo {author}
  {\bibfnamefont {C.~L.}\ \bibnamefont {Chien}},\ }\href
  {https://doi.org/10.1103/PhysRevB.94.020403} {\bibfield  {journal} {\bibinfo
  {journal} {Phys. Rev. B}\ }\textbf {\bibinfo {volume} {94}},\ \bibinfo
  {pages} {020403} (\bibinfo {year} {2016})}\BibitemShut {NoStop}%
\bibitem [{\citenamefont {Das}\ \emph {et~al.}(2017)\citenamefont {Das},
  \citenamefont {Schoemaker}, \citenamefont {van Wees},\ and\ \citenamefont
  {Vera-Marun}}]{FMSHEExp_Das}%
  \BibitemOpen
  \bibfield  {author} {\bibinfo {author} {\bibfnamefont {K.~S.}\ \bibnamefont
  {Das}}, \bibinfo {author} {\bibfnamefont {W.~Y.}\ \bibnamefont {Schoemaker}},
  \bibinfo {author} {\bibfnamefont {B.~J.}\ \bibnamefont {van Wees}},\ and\
  \bibinfo {author} {\bibfnamefont {I.~J.}\ \bibnamefont {Vera-Marun}},\ }\href
  {https://doi.org/10.1103/PhysRevB.96.220408} {\bibfield  {journal} {\bibinfo
  {journal} {Phys. Rev. B}\ }\textbf {\bibinfo {volume} {96}},\ \bibinfo
  {pages} {220408} (\bibinfo {year} {2017})}\BibitemShut {NoStop}%
\bibitem [{\citenamefont {Wang}\ \emph {et~al.}(2019)\citenamefont {Wang},
  \citenamefont {Wang}, \citenamefont {Amin}, \citenamefont {Wang},
  \citenamefont {Radhakrishnan}, \citenamefont {Davidson}, \citenamefont
  {Allen}, \citenamefont {Silva}, \citenamefont {Ohldag}, \citenamefont
  {Balzar}, \citenamefont {Zink}, \citenamefont {Haney}, \citenamefont {Xiao},
  \citenamefont {Cahill}, \citenamefont {Lorenz},\ and\ \citenamefont
  {Fan}}]{FMSHEExp_Wang}%
  \BibitemOpen
  \bibfield  {author} {\bibinfo {author} {\bibfnamefont {W.}~\bibnamefont
  {Wang}}, \bibinfo {author} {\bibfnamefont {T.}~\bibnamefont {Wang}}, \bibinfo
  {author} {\bibfnamefont {V.~P.}\ \bibnamefont {Amin}}, \bibinfo {author}
  {\bibfnamefont {Y.}~\bibnamefont {Wang}}, \bibinfo {author} {\bibfnamefont
  {A.}~\bibnamefont {Radhakrishnan}}, \bibinfo {author} {\bibfnamefont
  {A.}~\bibnamefont {Davidson}}, \bibinfo {author} {\bibfnamefont {S.~R.}\
  \bibnamefont {Allen}}, \bibinfo {author} {\bibfnamefont {T.~J.}\ \bibnamefont
  {Silva}}, \bibinfo {author} {\bibfnamefont {H.}~\bibnamefont {Ohldag}},
  \bibinfo {author} {\bibfnamefont {D.}~\bibnamefont {Balzar}}, \bibinfo
  {author} {\bibfnamefont {B.~L.}\ \bibnamefont {Zink}}, \bibinfo {author}
  {\bibfnamefont {P.~M.}\ \bibnamefont {Haney}}, \bibinfo {author}
  {\bibfnamefont {J.~Q.}\ \bibnamefont {Xiao}}, \bibinfo {author}
  {\bibfnamefont {D.~G.}\ \bibnamefont {Cahill}}, \bibinfo {author}
  {\bibfnamefont {V.~O.}\ \bibnamefont {Lorenz}},\ and\ \bibinfo {author}
  {\bibfnamefont {X.}~\bibnamefont {Fan}},\ }\href
  {https://doi.org/10.1038/s41565-019-0504-0} {\bibfield  {journal} {\bibinfo
  {journal} {Nature Nanotechnology}\ }\textbf {\bibinfo {volume} {14}},\
  \bibinfo {pages} {819} (\bibinfo {year} {2019})}\BibitemShut {NoStop}%
\bibitem [{\citenamefont {Soya}\ \emph {et~al.}(2023)\citenamefont {Soya},
  \citenamefont {Yamada}, \citenamefont {Hamaya},\ and\ \citenamefont
  {Ando}}]{FMSHEExp_Soya}%
  \BibitemOpen
  \bibfield  {author} {\bibinfo {author} {\bibfnamefont {N.}~\bibnamefont
  {Soya}}, \bibinfo {author} {\bibfnamefont {M.}~\bibnamefont {Yamada}},
  \bibinfo {author} {\bibfnamefont {K.}~\bibnamefont {Hamaya}},\ and\ \bibinfo
  {author} {\bibfnamefont {K.}~\bibnamefont {Ando}},\ }\href
  {https://doi.org/10.1103/PhysRevLett.131.076702} {\bibfield  {journal}
  {\bibinfo  {journal} {Phys. Rev. Lett.}\ }\textbf {\bibinfo {volume} {131}},\
  \bibinfo {pages} {076702} (\bibinfo {year} {2023})}\BibitemShut {NoStop}%
\bibitem [{\citenamefont {Lifshits}\ and\ \citenamefont
  {Dyakonov}(2009)}]{Lifshits_1}%
  \BibitemOpen
  \bibfield  {author} {\bibinfo {author} {\bibfnamefont {M.~B.}\ \bibnamefont
  {Lifshits}}\ and\ \bibinfo {author} {\bibfnamefont {M.~I.}\ \bibnamefont
  {Dyakonov}},\ }\href {https://doi.org/10.1103/PhysRevLett.103.186601}
  {\bibfield  {journal} {\bibinfo  {journal} {Phys. Rev. Lett.}\ }\textbf
  {\bibinfo {volume} {103}},\ \bibinfo {pages} {186601} (\bibinfo {year}
  {2009})}\BibitemShut {NoStop}%
\bibitem [{\citenamefont {Pauyac}\ \emph {et~al.}(2018)\citenamefont {Pauyac},
  \citenamefont {Chshiev}, \citenamefont {Manchon},\ and\ \citenamefont
  {Nikolaev}}]{Pauyac_1}%
  \BibitemOpen
  \bibfield  {author} {\bibinfo {author} {\bibfnamefont {C.~O.}\ \bibnamefont
  {Pauyac}}, \bibinfo {author} {\bibfnamefont {M.}~\bibnamefont {Chshiev}},
  \bibinfo {author} {\bibfnamefont {A.}~\bibnamefont {Manchon}},\ and\ \bibinfo
  {author} {\bibfnamefont {S.~A.}\ \bibnamefont {Nikolaev}},\ }\href
  {https://doi.org/10.1103/PhysRevLett.120.176802} {\bibfield  {journal}
  {\bibinfo  {journal} {Phys. Rev. Lett.}\ }\textbf {\bibinfo {volume} {120}},\
  \bibinfo {pages} {176802} (\bibinfo {year} {2018})}\BibitemShut {NoStop}%
\bibitem [{\citenamefont {Gupta}\ \emph {et~al.}(2024)\citenamefont {Gupta},
  \citenamefont {Park}, \citenamefont {Swain}, \citenamefont {Mishra},
  \citenamefont {Amin},\ and\ \citenamefont {Bedanta}}]{FMSHEExpTheory_Park}%
  \BibitemOpen
  \bibfield  {author} {\bibinfo {author} {\bibfnamefont {P.}~\bibnamefont
  {Gupta}}, \bibinfo {author} {\bibfnamefont {I.~J.}\ \bibnamefont {Park}},
  \bibinfo {author} {\bibfnamefont {A.}~\bibnamefont {Swain}}, \bibinfo
  {author} {\bibfnamefont {A.}~\bibnamefont {Mishra}}, \bibinfo {author}
  {\bibfnamefont {V.~P.}\ \bibnamefont {Amin}},\ and\ \bibinfo {author}
  {\bibfnamefont {S.}~\bibnamefont {Bedanta}},\ }\href
  {https://doi.org/10.1103/PhysRevB.109.014437} {\bibfield  {journal} {\bibinfo
   {journal} {Phys. Rev. B}\ }\textbf {\bibinfo {volume} {109}},\ \bibinfo
  {pages} {014437} (\bibinfo {year} {2024})}\BibitemShut {NoStop}%
\bibitem [{\citenamefont {Vansteenkiste}\ \emph {et~al.}(2014)\citenamefont
  {Vansteenkiste}, \citenamefont {Leliaert}, \citenamefont {Dvornik},
  \citenamefont {Helsen}, \citenamefont {Garcia-Sanchez},\ and\ \citenamefont
  {Van~Waeyenberge}}]{MuMax_8}%
  \BibitemOpen
  \bibfield  {author} {\bibinfo {author} {\bibfnamefont {A.}~\bibnamefont
  {Vansteenkiste}}, \bibinfo {author} {\bibfnamefont {J.}~\bibnamefont
  {Leliaert}}, \bibinfo {author} {\bibfnamefont {M.}~\bibnamefont {Dvornik}},
  \bibinfo {author} {\bibfnamefont {M.}~\bibnamefont {Helsen}}, \bibinfo
  {author} {\bibfnamefont {F.}~\bibnamefont {Garcia-Sanchez}},\ and\ \bibinfo
  {author} {\bibfnamefont {B.}~\bibnamefont {Van~Waeyenberge}},\ }\href
  {https://doi.org/10.1063/1.4899186} {\bibfield  {journal} {\bibinfo
  {journal} {AIP Advances}\ }\textbf {\bibinfo {volume} {4}},\ \bibinfo {pages}
  {107133} (\bibinfo {year} {2014})}\BibitemShut {NoStop}%
\bibitem [{\citenamefont {Landau}\ and\ \citenamefont
  {Lifshitz}(1935)}]{LLG_9}%
  \BibitemOpen
  \bibfield  {author} {\bibinfo {author} {\bibfnamefont {L.~D.}\ \bibnamefont
  {Landau}}\ and\ \bibinfo {author} {\bibfnamefont {E.~M.}\ \bibnamefont
  {Lifshitz}}\ }(\bibinfo {year} {1935})\BibitemShut {NoStop}%
\bibitem [{\citenamefont {Gilbert}(1955)}]{Gilbert_10}%
  \BibitemOpen
  \bibfield  {author} {\bibinfo {author} {\bibfnamefont {T.~L.}\ \bibnamefont
  {Gilbert}},\ }\href@noop {} {\bibfield  {journal} {\bibinfo  {journal}
  {Physical Review D}\ }\textbf {\bibinfo {volume} {100}},\ \bibinfo {pages}
  {1243} (\bibinfo {year} {1955})}\BibitemShut {NoStop}%
\bibitem [{\citenamefont {Finocchio}\ \emph {et~al.}(2011)\citenamefont
  {Finocchio}, \citenamefont {Krivorotov}, \citenamefont {Cheng}, \citenamefont
  {Torres},\ and\ \citenamefont {Azzerboni}}]{Finocchio_11}%
  \BibitemOpen
  \bibfield  {author} {\bibinfo {author} {\bibfnamefont {G.}~\bibnamefont
  {Finocchio}}, \bibinfo {author} {\bibfnamefont {I.~N.}\ \bibnamefont
  {Krivorotov}}, \bibinfo {author} {\bibfnamefont {X.}~\bibnamefont {Cheng}},
  \bibinfo {author} {\bibfnamefont {L.}~\bibnamefont {Torres}},\ and\ \bibinfo
  {author} {\bibfnamefont {B.}~\bibnamefont {Azzerboni}},\ }\href
  {https://doi.org/10.1103/PhysRevB.83.134402} {\bibfield  {journal} {\bibinfo
  {journal} {Phys. Rev. B}\ }\textbf {\bibinfo {volume} {83}},\ \bibinfo
  {pages} {134402} (\bibinfo {year} {2011})}\BibitemShut {NoStop}%
\bibitem [{\citenamefont {Leliaert}\ \emph {et~al.}(2017)\citenamefont
  {Leliaert}, \citenamefont {Mulkers}, \citenamefont {De~Clercq}, \citenamefont
  {Coene}, \citenamefont {Dvornik},\ and\ \citenamefont
  {Van~Waeyenberge}}]{leliaert_12}%
  \BibitemOpen
  \bibfield  {author} {\bibinfo {author} {\bibfnamefont {J.}~\bibnamefont
  {Leliaert}}, \bibinfo {author} {\bibfnamefont {J.}~\bibnamefont {Mulkers}},
  \bibinfo {author} {\bibfnamefont {J.}~\bibnamefont {De~Clercq}}, \bibinfo
  {author} {\bibfnamefont {A.}~\bibnamefont {Coene}}, \bibinfo {author}
  {\bibfnamefont {M.}~\bibnamefont {Dvornik}},\ and\ \bibinfo {author}
  {\bibfnamefont {B.}~\bibnamefont {Van~Waeyenberge}},\ }\href
  {https://doi.org/10.1063/1.5003957} {\bibfield  {journal} {\bibinfo
  {journal} {AIP Advances}\ }\textbf {\bibinfo {volume} {7}},\ \bibinfo {pages}
  {125010} (\bibinfo {year} {2017})}\BibitemShut {NoStop}%
\bibitem [{\citenamefont {Brown}(1963)}]{Brown_13}%
  \BibitemOpen
  \bibfield  {author} {\bibinfo {author} {\bibfnamefont {W.~F.}\ \bibnamefont
  {Brown}},\ }\href {https://doi.org/10.1103/PhysRev.130.1677} {\bibfield
  {journal} {\bibinfo  {journal} {Phys. Rev.}\ }\textbf {\bibinfo {volume}
  {130}},\ \bibinfo {pages} {1677} (\bibinfo {year} {1963})}\BibitemShut
  {NoStop}%
\bibitem [{\citenamefont {Belrhazi}\ \emph {et~al.}(2018)\citenamefont
  {Belrhazi}, \citenamefont {El~Hafidi},\ and\ \citenamefont
  {El~Hafidi}}]{currentDesnityRange}%
  \BibitemOpen
  \bibfield  {author} {\bibinfo {author} {\bibfnamefont {H.}~\bibnamefont
  {Belrhazi}}, \bibinfo {author} {\bibfnamefont {M.~Y.}\ \bibnamefont
  {El~Hafidi}},\ and\ \bibinfo {author} {\bibfnamefont {M.}~\bibnamefont
  {El~Hafidi}},\ }\href {https://doi.org/10.1007/s42452-018-0042-7} {\bibfield
  {journal} {\bibinfo  {journal} {SN Applied Sciences}\ }\textbf {\bibinfo
  {volume} {1}},\ \bibinfo {pages} {41} (\bibinfo {year} {2018})}\BibitemShut
  {NoStop}%
\bibitem [{\citenamefont {Coelho}\ \emph {et~al.}(2014)\citenamefont {Coelho},
  \citenamefont {Leitao}, \citenamefont {Antunes}, \citenamefont {Cardoso},\
  and\ \citenamefont {Freitas}}]{Coelho2014}%
  \BibitemOpen
  \bibfield  {author} {\bibinfo {author} {\bibfnamefont {P.}~\bibnamefont
  {Coelho}}, \bibinfo {author} {\bibfnamefont {D.~C.}\ \bibnamefont {Leitao}},
  \bibinfo {author} {\bibfnamefont {J.}~\bibnamefont {Antunes}}, \bibinfo
  {author} {\bibfnamefont {S.}~\bibnamefont {Cardoso}},\ and\ \bibinfo {author}
  {\bibfnamefont {P.~P.}\ \bibnamefont {Freitas}},\ }\bibfield  {journal}
  {\bibinfo  {journal} {IEEE Trans. Magn.}\ }\textbf {\bibinfo {volume} {50}},\
  \href {https://doi.org/10.1109/TMAG.2014.2321044} {10.1109/TMAG.2014.2321044}
  (\bibinfo {year} {2014})\BibitemShut {NoStop}%
\bibitem [{\citenamefont {Montoya}\ \emph {et~al.}(2023)\citenamefont
  {Montoya}, \citenamefont {Khan}, \citenamefont {Safranski}, \citenamefont
  {Smith},\ and\ \citenamefont {Krivorotov}}]{Montoya2023}%
  \BibitemOpen
  \bibfield  {author} {\bibinfo {author} {\bibfnamefont {E.~A.}\ \bibnamefont
  {Montoya}}, \bibinfo {author} {\bibfnamefont {A.}~\bibnamefont {Khan}},
  \bibinfo {author} {\bibfnamefont {C.}~\bibnamefont {Safranski}}, \bibinfo
  {author} {\bibfnamefont {A.}~\bibnamefont {Smith}},\ and\ \bibinfo {author}
  {\bibfnamefont {I.~N.}\ \bibnamefont {Krivorotov}},\ }\href
  {https://doi.org/10.1038/s42005-023-00935-1} {\bibfield  {journal} {\bibinfo
  {journal} {Commun. Phys.}\ }\textbf {\bibinfo {volume} {6}},\ \bibinfo
  {pages} {184} (\bibinfo {year} {2023})}\BibitemShut {NoStop}%
\bibitem [{\citenamefont {Takei}\ \emph {et~al.}(2017)\citenamefont {Takei},
  \citenamefont {Tserkovnyak},\ and\ \citenamefont {Mohseni}}]{Takei2017}%
  \BibitemOpen
  \bibfield  {author} {\bibinfo {author} {\bibfnamefont {S.}~\bibnamefont
  {Takei}}, \bibinfo {author} {\bibfnamefont {Y.}~\bibnamefont {Tserkovnyak}},\
  and\ \bibinfo {author} {\bibfnamefont {M.}~\bibnamefont {Mohseni}},\ }\href
  {https://doi.org/10.1103/PhysRevB.95.144402} {\bibfield  {journal} {\bibinfo
  {journal} {Phys. Rev. B}\ }\textbf {\bibinfo {volume} {95}},\ \bibinfo
  {pages} {144402} (\bibinfo {year} {2017})}\BibitemShut {NoStop}%
\bibitem [{\citenamefont {Schoen}\ \emph {et~al.}(2017)\citenamefont {Schoen},
  \citenamefont {Lucassen}, \citenamefont {Nembach}, \citenamefont {Koopmans},
  \citenamefont {Silva}, \citenamefont {Back},\ and\ \citenamefont
  {Shaw}}]{Schoen2017_damping}%
  \BibitemOpen
  \bibfield  {author} {\bibinfo {author} {\bibfnamefont {M.~A.~W.}\
  \bibnamefont {Schoen}}, \bibinfo {author} {\bibfnamefont {J.}~\bibnamefont
  {Lucassen}}, \bibinfo {author} {\bibfnamefont {H.~T.}\ \bibnamefont
  {Nembach}}, \bibinfo {author} {\bibfnamefont {B.}~\bibnamefont {Koopmans}},
  \bibinfo {author} {\bibfnamefont {T.~J.}\ \bibnamefont {Silva}}, \bibinfo
  {author} {\bibfnamefont {C.~H.}\ \bibnamefont {Back}},\ and\ \bibinfo
  {author} {\bibfnamefont {J.~M.}\ \bibnamefont {Shaw}},\ }\href
  {https://doi.org/10.1103/PhysRevB.95.134411} {\bibfield  {journal} {\bibinfo
  {journal} {Phys. Rev. B}\ }\textbf {\bibinfo {volume} {95}},\ \bibinfo
  {pages} {134411} (\bibinfo {year} {2017})}\BibitemShut {NoStop}%
\bibitem [{\citenamefont {Smith}\ \emph {et~al.}(2020)\citenamefont {Smith},
  \citenamefont {Rai}, \citenamefont {Lim}, \citenamefont {Hartnett},
  \citenamefont {Sapkota}, \citenamefont {Srivastava}, \citenamefont {Mewes},
  \citenamefont {Jiang}, \citenamefont {Clavel}, \citenamefont {Hudait},
  \citenamefont {Viehland}, \citenamefont {Heremans}, \citenamefont
  {Balachandran}, \citenamefont {Mewes},\ and\ \citenamefont
  {Emori}}]{Smith2020}%
  \BibitemOpen
  \bibfield  {author} {\bibinfo {author} {\bibfnamefont {D.~A.}\ \bibnamefont
  {Smith}}, \bibinfo {author} {\bibfnamefont {A.}~\bibnamefont {Rai}}, \bibinfo
  {author} {\bibfnamefont {Y.}~\bibnamefont {Lim}}, \bibinfo {author}
  {\bibfnamefont {T.}~\bibnamefont {Hartnett}}, \bibinfo {author}
  {\bibfnamefont {A.}~\bibnamefont {Sapkota}}, \bibinfo {author} {\bibfnamefont
  {A.}~\bibnamefont {Srivastava}}, \bibinfo {author} {\bibfnamefont
  {C.}~\bibnamefont {Mewes}}, \bibinfo {author} {\bibfnamefont
  {Z.}~\bibnamefont {Jiang}}, \bibinfo {author} {\bibfnamefont
  {M.}~\bibnamefont {Clavel}}, \bibinfo {author} {\bibfnamefont {M.~K.}\
  \bibnamefont {Hudait}}, \bibinfo {author} {\bibfnamefont {D.~D.}\
  \bibnamefont {Viehland}}, \bibinfo {author} {\bibfnamefont {J.~J.}\
  \bibnamefont {Heremans}}, \bibinfo {author} {\bibfnamefont {P.~V.}\
  \bibnamefont {Balachandran}}, \bibinfo {author} {\bibfnamefont
  {T.}~\bibnamefont {Mewes}},\ and\ \bibinfo {author} {\bibfnamefont
  {S.}~\bibnamefont {Emori}},\ }\href
  {https://doi.org/10.1103/PhysRevApplied.14.034042} {\bibfield  {journal}
  {\bibinfo  {journal} {Phys. Rev. Appl.}\ }\textbf {\bibinfo {volume} {14}},\
  \bibinfo {pages} {034042} (\bibinfo {year} {2020})}\BibitemShut {NoStop}%
\bibitem [{\citenamefont {Arora}\ \emph {et~al.}(2021)\citenamefont {Arora},
  \citenamefont {Delczeg-Czirjak}, \citenamefont {Riley}, \citenamefont
  {Silva}, \citenamefont {Nembach}, \citenamefont {Eriksson},\ and\
  \citenamefont {Shaw}}]{Arora2021}%
  \BibitemOpen
  \bibfield  {author} {\bibinfo {author} {\bibfnamefont {M.}~\bibnamefont
  {Arora}}, \bibinfo {author} {\bibfnamefont {E.~K.}\ \bibnamefont
  {Delczeg-Czirjak}}, \bibinfo {author} {\bibfnamefont {G.}~\bibnamefont
  {Riley}}, \bibinfo {author} {\bibfnamefont {T.~J.}\ \bibnamefont {Silva}},
  \bibinfo {author} {\bibfnamefont {H.~T.}\ \bibnamefont {Nembach}}, \bibinfo
  {author} {\bibfnamefont {O.}~\bibnamefont {Eriksson}},\ and\ \bibinfo
  {author} {\bibfnamefont {J.~M.}\ \bibnamefont {Shaw}},\ }\href
  {https://doi.org/10.1103/PhysRevApplied.15.054031} {\bibfield  {journal}
  {\bibinfo  {journal} {Phys. Rev. Applied}\ }\textbf {\bibinfo {volume}
  {15}},\ \bibinfo {pages} {054031} (\bibinfo {year} {2021})}\BibitemShut
  {NoStop}%
\bibitem [{\citenamefont {Maizel}\ \emph {et~al.}(2024)\citenamefont {Maizel},
  \citenamefont {Wu}, \citenamefont {Balakrishnan}, \citenamefont {Grutter},
  \citenamefont {Kinane}, \citenamefont {Caruana}, \citenamefont {Nakarmi},
  \citenamefont {Nepal}, \citenamefont {Smith}, \citenamefont {Lim},
  \citenamefont {Jones}, \citenamefont {Thomas}, \citenamefont {Zhao},
  \citenamefont {Michel}, \citenamefont {Mewes},\ and\ \citenamefont
  {Emori}}]{Maizel2024}%
  \BibitemOpen
  \bibfield  {author} {\bibinfo {author} {\bibfnamefont {R.~E.}\ \bibnamefont
  {Maizel}}, \bibinfo {author} {\bibfnamefont {S.}~\bibnamefont {Wu}}, \bibinfo
  {author} {\bibfnamefont {P.~P.}\ \bibnamefont {Balakrishnan}}, \bibinfo
  {author} {\bibfnamefont {A.~J.}\ \bibnamefont {Grutter}}, \bibinfo {author}
  {\bibfnamefont {C.~J.}\ \bibnamefont {Kinane}}, \bibinfo {author}
  {\bibfnamefont {A.~J.}\ \bibnamefont {Caruana}}, \bibinfo {author}
  {\bibfnamefont {P.}~\bibnamefont {Nakarmi}}, \bibinfo {author} {\bibfnamefont
  {B.}~\bibnamefont {Nepal}}, \bibinfo {author} {\bibfnamefont {D.~A.}\
  \bibnamefont {Smith}}, \bibinfo {author} {\bibfnamefont {Y.}~\bibnamefont
  {Lim}}, \bibinfo {author} {\bibfnamefont {J.~L.}\ \bibnamefont {Jones}},
  \bibinfo {author} {\bibfnamefont {W.~C.}\ \bibnamefont {Thomas}}, \bibinfo
  {author} {\bibfnamefont {J.}~\bibnamefont {Zhao}}, \bibinfo {author}
  {\bibfnamefont {F.~M.}\ \bibnamefont {Michel}}, \bibinfo {author}
  {\bibfnamefont {T.}~\bibnamefont {Mewes}},\ and\ \bibinfo {author}
  {\bibfnamefont {S.}~\bibnamefont {Emori}},\ }\href
  {https://arxiv.org/abs/2406.09874} {\bibfield  {journal} {\bibinfo  {journal}
  {arXiv:2406.09874}\ } (\bibinfo {year} {2024})}\BibitemShut {NoStop}%
\bibitem [{\citenamefont {Parkin}(1993)}]{Parkin1993}%
  \BibitemOpen
  \bibfield  {author} {\bibinfo {author} {\bibfnamefont {S.~S.~P.}\
  \bibnamefont {Parkin}},\ }\href {https://doi.org/10.1103/PhysRevLett.71.1641}
  {\bibfield  {journal} {\bibinfo  {journal} {Phys. Rev. Lett.}\ }\textbf
  {\bibinfo {volume} {71}},\ \bibinfo {pages} {1641} (\bibinfo {year}
  {1993})}\BibitemShut {NoStop}%
\bibitem [{\citenamefont {Swagten}\ \emph {et~al.}(1996)\citenamefont
  {Swagten}, \citenamefont {Strijkers}, \citenamefont {Bloemen}, \citenamefont
  {Willekens},\ and\ \citenamefont {de~Jonge}}]{Swagten1996}%
  \BibitemOpen
  \bibfield  {author} {\bibinfo {author} {\bibfnamefont {H.~J.~M.}\
  \bibnamefont {Swagten}}, \bibinfo {author} {\bibfnamefont {G.~J.}\
  \bibnamefont {Strijkers}}, \bibinfo {author} {\bibfnamefont {P.~J.~H.}\
  \bibnamefont {Bloemen}}, \bibinfo {author} {\bibfnamefont {M.~M.~H.}\
  \bibnamefont {Willekens}},\ and\ \bibinfo {author} {\bibfnamefont {W.~J.~M.}\
  \bibnamefont {de~Jonge}},\ }\href {https://doi.org/10.1103/PhysRevB.53.9108}
  {\bibfield  {journal} {\bibinfo  {journal} {Phys. Rev. B}\ }\textbf {\bibinfo
  {volume} {53}},\ \bibinfo {pages} {9108} (\bibinfo {year}
  {1996})}\BibitemShut {NoStop}%
\bibitem [{\citenamefont {Pai}\ \emph {et~al.}(2012)\citenamefont {Pai},
  \citenamefont {Liu}, \citenamefont {Li}, \citenamefont {Tseng}, \citenamefont
  {Ralph},\ and\ \citenamefont {Buhrman}}]{Pai2012}%
  \BibitemOpen
  \bibfield  {author} {\bibinfo {author} {\bibfnamefont {C.-F.}\ \bibnamefont
  {Pai}}, \bibinfo {author} {\bibfnamefont {L.}~\bibnamefont {Liu}}, \bibinfo
  {author} {\bibfnamefont {Y.}~\bibnamefont {Li}}, \bibinfo {author}
  {\bibfnamefont {H.~W.}\ \bibnamefont {Tseng}}, \bibinfo {author}
  {\bibfnamefont {D.~C.}\ \bibnamefont {Ralph}},\ and\ \bibinfo {author}
  {\bibfnamefont {R.~A.}\ \bibnamefont {Buhrman}},\ }\href
  {https://doi.org/10.1063/1.4753947} {\bibfield  {journal} {\bibinfo
  {journal} {Appl. Phys. Lett.}\ }\textbf {\bibinfo {volume} {101}},\ \bibinfo
  {pages} {122404} (\bibinfo {year} {2012})}\BibitemShut {NoStop}%
\bibitem [{\citenamefont {Demasius}\ \emph {et~al.}(2016)\citenamefont
  {Demasius}, \citenamefont {Phung}, \citenamefont {Zhang}, \citenamefont
  {Hughes}, \citenamefont {Yang}, \citenamefont {Kellock}, \citenamefont {Han},
  \citenamefont {Pushp},\ and\ \citenamefont {Parkin}}]{Demasius2016}%
  \BibitemOpen
  \bibfield  {author} {\bibinfo {author} {\bibfnamefont {K.-U.}\ \bibnamefont
  {Demasius}}, \bibinfo {author} {\bibfnamefont {T.}~\bibnamefont {Phung}},
  \bibinfo {author} {\bibfnamefont {W.}~\bibnamefont {Zhang}}, \bibinfo
  {author} {\bibfnamefont {B.~P.}\ \bibnamefont {Hughes}}, \bibinfo {author}
  {\bibfnamefont {S.-H.}\ \bibnamefont {Yang}}, \bibinfo {author}
  {\bibfnamefont {A.}~\bibnamefont {Kellock}}, \bibinfo {author} {\bibfnamefont
  {W.}~\bibnamefont {Han}}, \bibinfo {author} {\bibfnamefont {A.}~\bibnamefont
  {Pushp}},\ and\ \bibinfo {author} {\bibfnamefont {S.~S.~P.}\ \bibnamefont
  {Parkin}},\ }\href {https://doi.org/10.1038/ncomms10644} {\bibfield
  {journal} {\bibinfo  {journal} {Nat. Commun.}\ }\textbf {\bibinfo {volume}
  {7}},\ \bibinfo {pages} {10644} (\bibinfo {year} {2016})}\BibitemShut
  {NoStop}%
\bibitem [{\citenamefont {Hrabec}\ \emph {et~al.}(2016)\citenamefont {Hrabec},
  \citenamefont {Gonçalves}, \citenamefont {Spencer}, \citenamefont
  {Arenholz}, \citenamefont {N'Diaye}, \citenamefont {Stamps},\ and\
  \citenamefont {Marrows}}]{Hrabec2016}%
  \BibitemOpen
  \bibfield  {author} {\bibinfo {author} {\bibfnamefont {A.}~\bibnamefont
  {Hrabec}}, \bibinfo {author} {\bibfnamefont {F.~J.~T.}\ \bibnamefont
  {Gonçalves}}, \bibinfo {author} {\bibfnamefont {C.~S.}\ \bibnamefont
  {Spencer}}, \bibinfo {author} {\bibfnamefont {E.}~\bibnamefont {Arenholz}},
  \bibinfo {author} {\bibfnamefont {A.~T.}\ \bibnamefont {N'Diaye}}, \bibinfo
  {author} {\bibfnamefont {R.~L.}\ \bibnamefont {Stamps}},\ and\ \bibinfo
  {author} {\bibfnamefont {C.~H.}\ \bibnamefont {Marrows}},\ }\href
  {https://doi.org/10.1103/PhysRevB.93.014432} {\bibfield  {journal} {\bibinfo
  {journal} {Phys. Rev. B}\ }\textbf {\bibinfo {volume} {93}},\ \bibinfo
  {pages} {014432} (\bibinfo {year} {2016})}\BibitemShut {NoStop}%
\end{thebibliography}%
\end{document}